\documentclass[12pt]{iopart}

\usepackage{amssymb}
\usepackage{graphicx}
\usepackage{dcolumn}
\usepackage{bm}
\usepackage{psfrag}
\usepackage{bbold}
\usepackage{amsthm}
\usepackage{iopams}
\usepackage{color}

\newcommand{\decay}{\Gamma}
\newcommand{\deph}{\gamma}
\newcommand{\be}{\begin{equation}}
\newcommand{\ee}{\end{equation}}
\newcommand{\sysb}{\left\{\begin{array}}
\newcommand{\syse}{\end{array}\right.}
\newcommand{\baa}{\begin{array}}
\newcommand{\eaa}{\end{array}}

\newcommand{\mal}{\mathcal}

\newcommand{\lt}{\left(}
\newcommand{\rt}{\right)}
\newcommand{\lqq}{\left[}
\newcommand{\rqq}{\right]}

\newcommand{\id}{\mathbb{1}}
\newcommand{\eval}[1]{\left.\right|_{ #1 }}

\newcommand{\matb}{\left(\begin{array}}
\newcommand{\mate}{\end{array}\right)}

\newcommand{\sx}{\sigma^x}
\newcommand{\sy}{\sigma^y}
\newcommand{\sz}{\sigma^z}
\newcommand{\skx}[1]{\sigma^x_{#1}}

\newcommand{\lan}{\left\langle}
\newcommand{\ran}{\right\rangle}
\newcommand{\ket}[1]{\left| #1 \ran}
\newcommand{\bra}[1]{\lan #1 \right|}

\newcommand{\outr}[2]{\left| #1 \ran \!\lan #2 \right|}
\newcommand{\proj}[1]{\ket{#1} \bra{#1}}
\newcommand{\comm}[2]{\left[ #1, #2 \right]}
\newcommand{\acomm}[2]{\left\{ #1, #2 \right\}}

\newcommand{\av}[1]{\left\langle  #1  \right\rangle}

\newcommand{\cosa}[1]{\cos \left(  #1 \right)}

\newcommand{\nol}{\nonumber \\}

\newcommand{\reff}[1]{(\ref{#1})}

\newcommand{\prodl}[2]{\prod\limits_{#1}^{#2}}
\newcommand{\suml}[2]{\sum\limits_{#1}^{#2}}

\newcommand{\liml}[1]{\lim\limits_{#1}}

\newcommand{\N}{\mathbb{N}}

\newcommand{\comma}{\quad , \quad}

\newcommand{\LO}{\mal{L}_0}
\newcommand{\LI}{\mal{L}_1}

\newcommand{\ha}{\frac{1}{2}}

\newcommand{\wt}{\widetilde}

\begin{document}

\title{Effective dynamics of strongly dissipative Rydberg gases}

\author{M Marcuzzi, J Schick, B Olmos and I Lesanovsky}
\address{School of Physics and Astronomy, The University of Nottingham, Nottingham, NG7 2RD, United Kingdom}

\begin{abstract}
  We investigate the evolution of interacting Rydberg gases in the limit of strong noise and dissipation. Starting from a description in terms of a Markovian quantum master equation we derive effective equations of motion that govern the dynamics on a ``coarse-grained'' timescale where fast dissipative degrees of freedom have been adiabatically eliminated. Specifically, we consider two scenarios which are of relevance for current theoretical and experimental studies --- Rydberg atoms in a two-level (spin) approximation subject to strong dephasing noise as well as Rydberg atoms under so-called electromagnetically induced transparency (EIT) conditions and fast radiative decay. In the former case we find that the effective dynamics is described by classical rate equations up to second order in an appropriate perturbative expansion. This drastically reduces the computational complexity of numerical simulations in comparison to the full quantum master equation. When accounting for the fourth order correction in this expansion, however, we find that the resulting equation breaks the preservation of positivity and thus cannot be interpreted as a proper classical master rate equation. In the EIT system we find that the expansion up to second order retains information not only on the ``classical'' observables, but also on some quantum coherences. Nevertheless, this perturbative treatment still achieves a non-trivial reduction of complexity with respect to the original problem.

  \noindent{\it Keywords\/}:
\end{abstract}
\pacs{...}
\maketitle

\section{Introduction}

Gases of interacting highly excited Rydberg atoms are becoming an increasingly popular theoretical and experimental platform for the investigation of the physics of strongly interacting many-body systems \cite{Gallagher84, Low12}. The main distinction between these systems and ``traditional'' ones restricted to low-lying excited states lies in the huge enhancement in the interaction between the atoms, which can be several orders of magnitude stronger for typical experimental parameters. Indeed, two atoms in a Rydberg state usually experience extremely strong dipole-dipole or van der Waals forces (see e.g., \cite{Saffman10-2}). These in turn considerably affect both the static \cite{Weimer08, Low09, Weimer10-2, Schauss12, Levi14} and dynamic \cite{Schachenmayer10, Pohl10, Bettelli13, Schaub14, Lesanovsky14, Hoening14} properties of the system.

The rather long lifetime of these Rydberg states allows for probing of the coherent evolution of these many-body systems up to relatively long time-scales. Here, however, we will focus on situations (currently studied with great interest) where accounting for dissipative processes leads to interesting changes in both the dynamics and the stationary properties. These processes emerge due to the coupling of the system to external degrees of freedom, which produce e.g. decay via spontaneous emission (fluorescence) and noise-induced loss of coherence (dephasing). The evolution is typically well described in terms of a quantum master equation with Markovian noise (an overview on methods for treating the non-Markovian case as well can be found in \cite{Breuer99}). While such a modelling is certainly among the simplest descriptions of an open quantum system it still poses severe challenges when trying to conduct a numerical treatment for large system sizes $N$. This is due to the very fast increase in the dimension of the many-body Liouville space ($b^{2N}$ with $b$ being the dimension of the single-particle Hilbert space, e.g., $b = 2$ for an Ising spin). Some procedures to address this issue have been developed which divide the system into two parts, a subsystem of interest and an environment which is traced away \cite{Hartmann14, Breuer_book}.

A somewhat different framework arises when it is possible to identify degrees of freedom that evolve on vastly different timescales. Adiabatically eliminating fast-evolving ones might then allow the derivation of an effective equation of motion for the remaining slow degrees of freedom that portrays a reduced complexity and might be amenable to numerical treatment (see \ref{app:general} or \cite{Hartmann14} for a general description of the method, based on the Nakajima-Zwanzig projection formalism \cite{Nakajima58, Zwanzig60}). One of the first works to apply an idea along those lines to Rydberg systems was that of Ates et al. \cite{Ates06,Ates07-2}, where it was shown that --- in an appropriate limit --- properties of the stationary state of an interacting Rydberg gas can be extracted via a classical Monte-Carlo method. This made the simulation of stationary properties of large scale Rydberg systems feasible and several recent works employ variations of the same method \cite{Hoening14,Heeg12,Hoening13,Gaerttner13,Petrosyan13,Petrosyan13-1,Gaerttner14}.

Besides studying the properties of the stationary state there is great current interest in the understanding of the out-of-equilibrium dynamics of these interacting Rydberg gases. Here, effective equations of motion that describe the systems' dynamics in terms of a classical rate equation have been put forward and used by several authors both in purely theoretical works \cite{Hoening14,Lesanovsky13,Schoenleber14} and to model actual experimental data \cite{Schempp14,Urvoy14}.

\begin{figure}
\center
  \includegraphics[width=0.8\columnwidth]{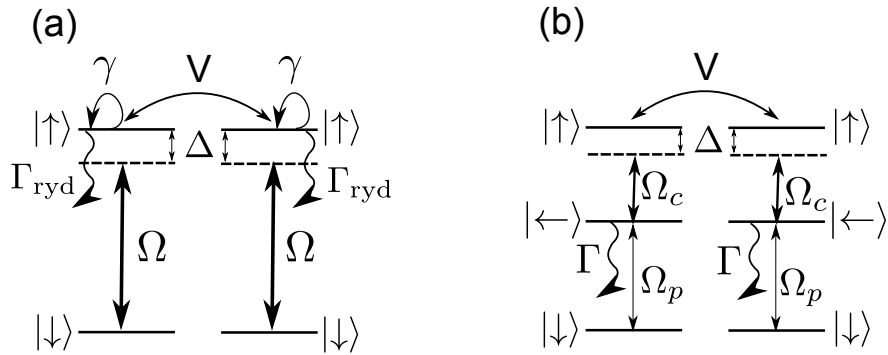}
  \caption{Atomic level schemes considered in this paper. Two atoms effectively only interact when they are both excited to the Rydberg state $\left|\uparrow\right>$; here we denote with $V$ the strength of such an interaction. Most of the current experiments can be modelled by the following two descriptions: \textbf{(a):} Two-level atoms driven by a laser with Rabi frequency $\Omega$ and detuning $\Delta$. The main dissipation mechanism we consider here is dephasing (at rate $\gamma$) of superpositions between the states $\left|\uparrow\right>$ and $\left|\downarrow\right>$. We will also take into account the decay from $\left|\uparrow\right>$ to $\left|\downarrow\right>$ with rate $\Gamma_\mathrm{ryd}$, which is usually small compared to $\gamma$ and thus will be treated perturbatively. \textbf{(b):} Three-level atoms in an EIT configuration, i.e. where the excitation of the Rydberg states is performed via a transition between the ground state $\left|\downarrow\right>$ and the intermediate state $\left|\leftarrow\right>$. Within the time-scales of interest, spontaneous decay processes are assumed to be relevant only for the intermediate state, whose inverse lifetime is $\Gamma$.}
\label{fig:schemes}
\end{figure}

The purpose of this paper is to give a detailed account of the derivation of effective equations of motion that describe the many-body dynamics of interacting Rydberg gases in the limit of strong dissipation. Specifically, we will discuss the two scenarios depicted in figure \ref{fig:schemes} that are directly relevant to current experiments:

The first one is that of a Rydberg gas in which atoms are modelled by coherently driven and interacting two-level systems. Here dissipation is present in the form of dephasing noise that quickly destroys coherent superpositions between the two states. This corresponds approximately to the experimental situations discussed in Refs.~\cite{Schempp14,Urvoy14}. We show that in the limit of strong dephasing the dynamics of the interacting two-level systems is described --- up to second order in the relevant perturbative expansion --- by a classical rate equation; the corresponding stochastic process is described by single spin flips subject to kinetic constraints \cite{Ritort2003}. We then proceed further and calculate the fourth-order corrections, showing that they result in new processes, such as simultaneous two-spin flips. However, it turns out that, unlike for the second order case, there are domains in the space of physical parameters for which some of the ``rates'' become negative, thereby breaking the conservation of positivity. Hence, a standard treatment in terms of a classical stochastic dynamics is not always possible. Yet, the perturbative expansion is formally correct and our numerical analysis shows that said breakdown only affects the initial stage of the dynamics, whereas for long times a good agreement with numerically exact data is still found. We conclude the discussion by including (radiative) decay from the upper to the lower atomic level in the rate equation treatment.

The second scenario we are considering is that of Rydberg gases under so-called electromagnetically induced transparency (EIT) conditions. In this regime, which has been studied experimentally in Refs.~\cite{Pritchard10,Adams13,Schempp10,Schwarzkopf11,Schwarzkopf13,Peyronel12,Maxwell13}, atoms are modelled by coherently driven and interacting three-level systems. Dissipation enters through a fast (radiative) decay of the middle level to the lower one. In this case, the fast dissipative dynamics is of a different nature and does not necessarily project onto a classical spin configuration. Therefore, information on some quantum coherences must be retained. Despite the fact that one does not gain as simple a description as in the case above, the resulting reduced equation is still in Lindblad form and offers therefore a simplified alternative to the study of the one acting on the whole Hilbert space. In particular, the interatomic interaction needs to be taken into account coherently, while the elimination of the fast decay leads to an unusual form of effective dissipation that drives the system into coherent superposition states. We conclude the discussion of the three-level system by considering the limit of infinitely strong nearest-neighbour interaction, for which we find an effective purely-dissipative quantum dynamics that is reminiscent of that of quantum analogues of kinetically constrained spin models \cite{Olmos14}.

The paper is organized such that the central results and conclusions are presented in the main text, whilst a more detailed formal derivation is provided in the final appendices to which we refer in the appropriate sections.

\section{Two-level Rydberg atoms in the presence of strong dephasing noise}\label{sec:dephasing}

We consider here a gas of $N$ atoms with two relevant internal levels as shown in figure \ref{fig:schemes}(a). We assume that the timescale of the external motion of the atoms is much larger than the one in which the electronic dynamics takes place. This ``frozen gas'' picture has been shown to be adequate in a vast number of theoretical and experimental works. The ground state $\left|\downarrow\right\rangle$ and the Rydberg state $\left|\uparrow\right\rangle$ are coupled by a laser with Rabi frequency $\Omega$ and detuning $\Delta$. In order to avoid having an explicit angular dependence of the interaction, the excited state is typically chosen to have spherical symmetry, i.e., to be an $S$ orbital. Sharing the same parity of the ground state, however, it is not possible to reach it via a single dipole transition; in practice, this excitation must be achieved by means of a two-photon process via a far off-resonant excitation of an intermediate state. In this case the two-level approximation is adequate. Later we will, however, also account for the case of near-resonant excitation of such an intermediate level, which makes it necessary to include it as well in the description.

When two atoms ($k$ and $m$) are simultaneously excited to the Rydberg state they interact due to the electrostatic coupling of the respective (permanent or induced) dipole moments. The strength of this interaction $V_{km}$ is therefore of the form \cite{Saffman10-2}
\begin{equation*}
	V_{km}=\frac{C_p}{\left| \mathbf{r}_k-\mathbf{r}_m \right|^p},
\end{equation*}
where $\mathbf{r}_k$ denotes the position of the $k$-th atom, $C_p$ is the dispersion coefficient and $p$ characterizes the interaction type: $p = 3$ stands for dipole-dipole and $p=6$ for van der Waals forces. The dynamics of the density matrix $\rho$ of the system is described by a quantum master equation
\begin{equation*}
\dot{\rho}=-i\comm{H}{\rho}+\mal{D} \rho.
\end{equation*}
The first term describes the coherent evolution of the system which (within the rotating wave approximation) is governed by the many-body Hamiltonian $H=H_0+H_1$ where
\begin{equation*}
	\sysb{l}
	H_0=\Delta \suml{k}{} n_k +  \ha \suml{k \neq m}{} V_{km} n_k n_m\\
	H_1 = \Omega \suml{k}{} \sigma_k^x,
    \syse
\end{equation*}
with the operators $n_k=\outr{\uparrow_k}{\uparrow_k}$ and $\skx{k}=\outr{\downarrow_k}{\uparrow_k}+\outr{\uparrow_k}{\downarrow_k}$.

The generator of the dissipative dynamics is modeled in terms of a dissipator in Lindblad form which in case of the dephasing noise considered here is given by
\be
	\mal{D} \rho = \deph \suml{k}{}\lt n_k \rho n_k - \frac{1}{2} \acomm{n_k}{\rho} \rt.
	\label{eq:dissipator}
\ee
with $\gamma$ being the dephasing rate. Note, that this relies on the assumption that the noise can be considered white and spatially uncorrelated, i.e., acting independently on each atom. In practice, this is not always the case: for instance, dephasing noise can be introduced by fluctuations in the laser fields with a finite correlation length; if the typical interatomic distance is smaller than this correlation length, the noise experienced by nearby atoms will be spatially-correlated. Nevertheless, to consider independent fluctuations represents a reasonable approach for dilute Rydberg ensembles and, moreover, recent experimental work \cite{Schempp14,Urvoy14,Raitzsch09} suggests that this approximation captures the essential physics of the problem.

\subsection{Second order effective evolution}
We are now interested in deriving an effective equation of motion for the system in the regime where the dephasing rate is large or more precisely $\deph\gg\Omega$. In this limit coherent superpositions of the local atomic states $\left|\downarrow\right\rangle$ and $\left|\uparrow\right\rangle$ will dephase exponentially fast on timescales of the order of $\gamma^{-1}$. On timescales longer than this dephasing time the density matrix will therefore no longer show coherences and we can thus describe the system's state by a reduced density matrix $\mu$ which includes only the diagonal elements of $\rho$, i.e., only the probabilities of the classical spin configurations (direct products of the form $\ket{\cdots \uparrow \uparrow \downarrow \uparrow \cdots}$). Note that, because of this, the only observables which can be calculated within this scheme are the diagonal ones, i.e., those which can be written as combinations of $n_k$-s and the identity. A more formal version of the arguments above is given in \ref{app:dephasing}.

The coherent flipping induced by the laser provides a much slower dynamics which, due to the fast action of the dephasing, can be effectively projected onto the stationary subspace of the dephasing and accounted for in a perturbative expansion in powers of $\Omega$. Accordingly, the effective equation of motion can be cast in the form
\be
\dot{\mu}=\sum_{\alpha=1}^\infty \mathcal{L}^{(\alpha)} \mu,
\label{eq:muexp}
\ee
where $\mathcal{L}^{(\alpha)}$ is the evolution operator of order $\Omega^\alpha$. All odd terms of this series identically vanish, hence the first non-vanishing term is of second order, i.e. $\mathcal{L}^{(2)}$. The corresponding truncated evolution of $\mu$ at this level reads
\be \label{eq:M2}
\dot{\mu}=\mathcal{L}^{(2)} \mu=\sum_k \Gamma_k \left(\sigma_k^x\mu\sigma_k^x-\mu\right)
\ee
with
\be
 \Gamma_k=\frac{\Omega^2\deph}{\lt\frac{\deph}{2}\rt^2 +\lt\Delta + \sum_{q\neq k}V_{kq}n_q\rt^2},
 \label{eq:Gammak}
\ee
as was derived in \cite{Lesanovsky13}. Since $\mu$ is the diagonal matrix of probabilities associated to classical spin configurations, \reff{eq:M2} describes a continuous-time stochastic process which flips the $k$-th spin with operator-valued rate $\Gamma_k$. More precisely, the rate depends via the interaction term on the configurations of all sites but the $k$-th one. The expression \reff{eq:M2} is therefore equivalent to a \emph{kinetically-constrained} rate equation \cite{Ritort2003, JPG1, JPG2}, i.e., it can be regarded in terms of a trivial evolution (flipping one spin at a time) subject to a non-trivial constraint (here the number of excitations present in the neighbourhood). This can be more clearly understood by introducing the probability vector $\mathbf{v}=\mathrm{diag}(\mu)$ which evolves according to
\begin{equation*}
\dot{\mathbf{v}}=\sum_k\Gamma_k\left[\sigma_k^+-(1-n_k) \right]\mathbf{v}+\sum_k\Gamma_k\left[\sigma_k^- - n_k\right]\mathbf{v}.
\end{equation*}
where $\sigma^+_k=\outr{\uparrow_k}{\downarrow_k}$ and $\sigma^-_k=\outr{\downarrow_k}{\uparrow_k}$. Here each term describes the incoherent state change of an atom from $\left|\downarrow\right\rangle$ to $\left|\uparrow\right\rangle$, and viceversa, with rate $\Gamma_k$. This representation has the distinct advantage of allowing for numerical investigations of large scale systems by virtue of classical Monte-Carlo simulations. This fact has been exploited in a number of recent works \cite{Hoening14,Lesanovsky13,Schoenleber14}.

\begin{figure}
\includegraphics[width=0.9\columnwidth]{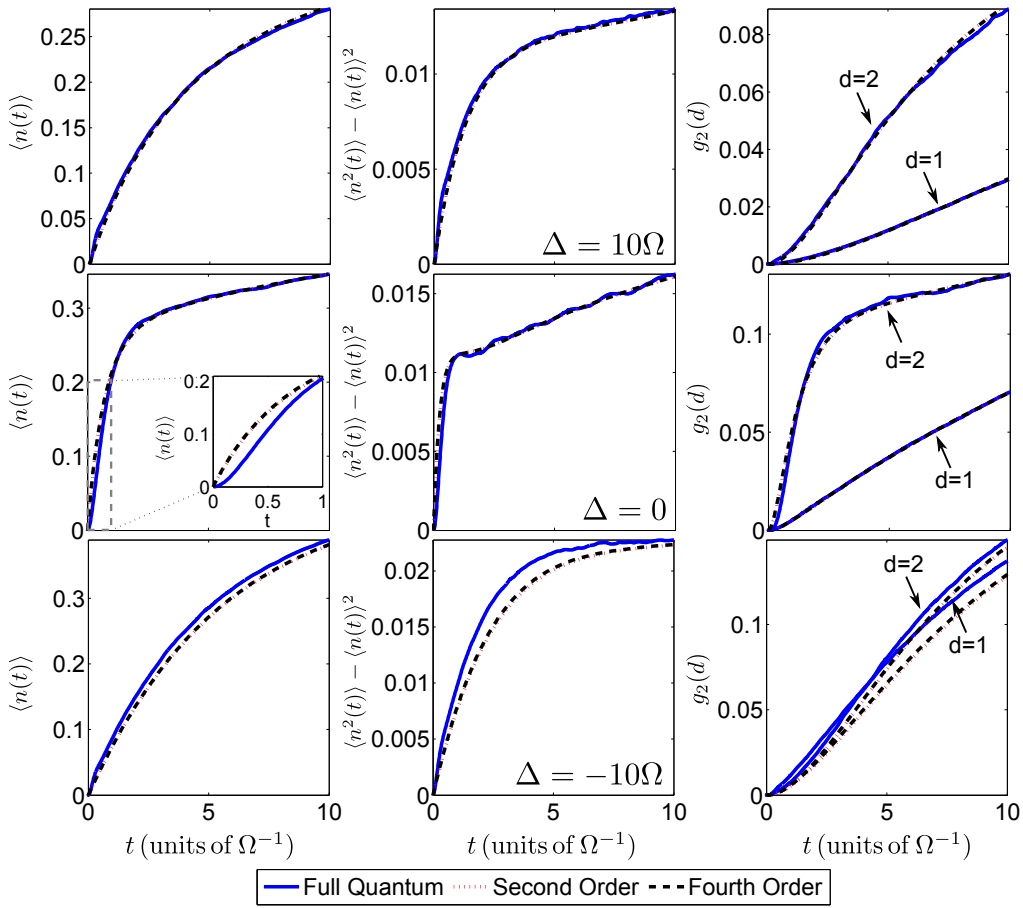}
\caption{Time evolution of the density of excitations $\langle n\rangle$, its associated fluctuations $\langle n^2\rangle-\langle n\rangle^2$ and the density-density correlations $g_2(d)$ for nearest ($d=1$) and next-nearest neighbours ($d=2$). In all cases the initial state is the one without excitations $\otimes_k \ket{\downarrow_k}$. We compare the results obtained from the Quantum Jump Monte-Carlo simulation of the full quantum system and the numerically exact integration of the effective Master equation obtained up to second [given by \reff{eq:M2}] and fourth order [adding the contribution given by \reff{eq:M4}] for $N=9$ atoms. The parameters used in the simulations shown are $\Delta=-10\Omega,0,10\Omega$, $\gamma=10\Omega$ and $V=10\Omega$.}\label{fig:dephasing}
\end{figure}

In order to assess the validity of this effective description, we have computed the evolution of small systems (up to $N=9$ atoms) accounting for both the full quantum master equation (numerically simulated via Quantum Jump Monte-Carlo \cite{Dalibard92,Molmer93}) and the resulting classical second order equation [numerically exact integration of \reff{eq:M2}]. For our simulations we have considered the atoms trapped in a one dimensional chain with lattice constant $a$, with one atom per site, periodic boundary conditions and van der Waals interaction. We have chosen a value for the dephasing compatible with the perturbative requirement, namely $\gamma=10\Omega$, and fixed the value of the nearest neighbour interaction to $V=C_6/a^6=10\Omega$. In figure \ref{fig:dephasing} we show for a system of $N=9$ atoms the short-time evolution of the mean density of excitations $\langle n\rangle=\sum_k\langle n_k\rangle/N$, its fluctuations $\langle n^2\rangle-\langle n\rangle^2$ and the density-density correlations
\begin{equation*}
g_2(d)=\frac{1}{N}\sum_k\langle n_kn_{k+d}\rangle,
\end{equation*}
between nearest ($d=1$) and next-nearest ($d=2$) neighbours, with $d$ being the distance in units of $a$. We choose to focus on the short time behaviour as here the difference between the exact and effective dynamics becomes most visible. For longer times, both dynamics reach the same stationary state, which is completely mixed, i.e., $\mu$ and $\rho$ become proportional to the identity. This is a consequence of the fact that the dissipator (\ref{eq:dissipator}) is constructed solely upon Hermitian jump operators $n_k$.

In general there is good agreement between the approximate and exact dynamics for the chosen parameter sets and observables. The initial increase in the density of excitations for very short times (smaller than $\Omega^{-1}$) is not expected to be well-captured by the approximation introduced above. Considering that the initial state is the completely polarised one $\bigotimes_k \ket{\downarrow_k}$, which belongs to the stationary space of the dephasing, in the full quantum problem the early stage of the dynamics is approximately driven by the coherent part and thus must be reversible. Therefore the density can only start from $0$ with vanishing slope, so that the initial increase of the excitation density is always proportional to (at least) $t^2$. This behavior cannot be captured by the classical rate equation \reff{eq:M2}. This can be understood by determining the equation of motion of the density of excitations $n_k$ at site $k$, which reads
\begin{equation*}
	\av{\dot{n}_k} = \Tr{\left(n_k \dot{\mu}\right)} =  \av{\Gamma_k\lt   1 - 2 n_k \rt}.
\end{equation*}
Due to the factorised nature of the initial state, $\av{\Gamma_k n_k} = \av{\Gamma_k} \av{n_k}$. Hence, starting from $\av{n_k}=0$, we observe that the gradient at $t=0$ is indeed different from $0$ and the density's initial increase is proportional to $t$ (see e.g. \cite{Urvoy14}). This difference is most obvious in the resonant case $\Delta=0$, where a magnified view of the very short-time regime is shown (see second row in figure \ref{fig:dephasing}).

For times beyond $\Omega^{-1}$ one can still observe small amplitude oscillations of the numerically exact solution; these are due to the dampening effects of the coherences and are thus not captured by the approximate dynamics. Compatibly with our assumptions, they  become less and less pronounced as $\gamma$ increases and off-diagonal terms are quickly damped out. Within the parameter regime analysed here, it also becomes apparent that the agreement is enhanced for positive values of the detuning. At an intuitive level, this can be related to the form of the rates \reff{eq:Gammak}, which can be interpreted as the effective perturbative parameters. It is clear that, for negative $\Delta$, the effects of the detuning and the interaction are competing and one can generally obtain rates which are smaller than in the case of positive detuning.

\subsection{Fourth order corrections}
The next non-vanishing order in the perturbative expansion \reff{eq:muexp} of the effective master equation is the fourth one ($\propto \Omega^4$). Its structure is considerably more involved than the second order contribution. It can be divided into five terms:
\be
	\mathcal{L}^{(4)} \mu    =    \Omega^4\sum_{k\neq m} \lt G^{km}_1 + G^{km}_2 + G^{km}_3 + G^{km}_4 \rt  \mu + \Omega^4 \sum_k F^{k} \mu
	\label{eq:M4}
\ee
where each superoperator ``$G$'' can be represented in general as
\begin{eqnarray}
	\label{eq:G}
  G_i^{km} \mu  &=&  R_i^{km} \sx_m  {R'}_i^{km} \sx_k \,  \mu  \, \sx_k  \, \sx_m - R_i^{km} \sx_m {R'}_i^{km}  \, \mu \, \, \sx_m  \nol
  &&  -  R_i^{km} {R'}_i^{km} (\sigma_k^x \,  \mu    \,\sigma_k^x - \mu) ,
\end{eqnarray}
with $R_i$ and $R'_i$ being hermitian operator-valued coefficients. Analogously to the second order rates above, the structure of these coefficients is diagonal, in the sense that they can be written as non-linear combinations of the local density operators $n_q$. Their specific form is, however,  much more complicated in this case:
\begin{equation*}
\sysb{l}
    {R}_1^{km} =\id \\[2mm]
	{R'}_1^{km} = 2\Re \lqq\lt {\Gamma^m_1}^\dagger {\Gamma_2^{km}}^\dagger + \Gamma^m_1 \Gamma_{3}^{km}  \rt {\Gamma_1^k}^\dagger\rqq\\[2mm]
    {R}_2^{km} =\Re\lt\Gamma_1^k\rt\\[2mm]
	{R'}_2^{km} =2\Re\lqq{\Gamma_1^k}^\dagger \lt {\Gamma_2^{km}}^\dagger +\Gamma_3^{km} \rt\rqq\\[2mm]
    {R}_3^{km} =-\Im\lt\Gamma_1^k\rt\\[2mm]
	{R'}_3^{km} =-2\Im\lqq{\Gamma_1^k}^\dagger \lt {\Gamma_2^{km}}^\dagger +\Gamma_3^{km} \rt\rqq\\[2mm]
    {R}_4^{km} = - 2   \lqq  \Re \lt \Gamma_1^m  \rt \rqq^2  + 2 \lqq \Im \lt \Gamma_1^m  \rt \rqq^2 \\[2mm] 	 		
	{R'}_4^{km} = 2\Re\lt\Gamma_1^k\rt,
\syse
\end{equation*}
where
\begin{equation*}
	\sysb{l}
		\Gamma_1^k = \frac{1}{ \frac{\gamma}{2}  + i \lqq  \Delta + \sum_{q\neq k} V_{kq} n_q  \rqq   }\\[3mm]
		\Gamma_{2}^{km} = \frac{1}{\gamma  + i \lqq 2\Delta + \sum_{q \neq k,m}  V_{kq} n_q + \sum_{q \neq k,m}  V_{mq} n_q   + V_{km} \rqq}\\[3mm]
		\Gamma_{3}^{km} = \frac{1}{\gamma  + i \lqq  \sum_{q \neq k,m}  \lt V_{mq} - V_{kq} \rt n_q  \rqq}.
	\syse
\end{equation*}
Note that the preservation of the trace is here ensured by the fact that every ${R'}^{km}_i$ except ${R'}^{km}_1$ commutes with $\sx_k$, whilst ${R}^{km}_{1,4}$ commute with $\sx_m$ and ${R}^{km}_{2,3}$ with $\sx_k$. The last term of \reff{eq:M4} is instead of the form
\begin{equation}\label{eq:F}
  F^k \mu   =   \beta_k \left(\sigma_k^x\mu\sigma_k^x-\mu\right),
\end{equation}
where $\beta_k$ is again an operator-valued rate (i.e., a kinetic constraint) commuting with $\sx_k$, which reads
\begin{equation*}
    \beta_k=  8 \lqq\Re \lt \Gamma_1^k  \rt\rqq^2  \left\{  \lqq\Re\lt\Gamma_1^k\rt\rqq^2- \lqq\Im\lt\Gamma_1^k\rt\rqq^2 \right\}.
\end{equation*}
A detailed derivation of the fourth order contribution to the perturbative expansion \reff{eq:M4} and the specific forms of the rates are given in \ref{app:dephasing}.

We now wish to give a stochastic interpretation to the terms resulting from this fourth order expansion. The action of the ``$F$'' superoperator \reff{eq:F} displays the same structure as the second order terms \reff{eq:M2}. Thus, it represents simply a perturbative correction of order $\Omega^4$ to these processes. The action of the ``$G$'' superoperators \reff{eq:G} is more involved. Here, we start by separating terms that lead to a single spin-flip from the ones that lead to two correlated spin flips. The former constitute an additional fourth-order correction to \reff{eq:M2}, whereas the latter introduce novel dynamical processes. Collecting all terms up to fourth order, the effective equation for the probability vector $\mathbf{v}$ reads now
\begin{eqnarray}\label{eq:upto4}
\dot{\mathbf{v}}&=&\sum_k\left[\sigma_k^+-(1-n_k)  \right]\Gamma_k^{\mathrm{s}}\mathbf{v}+\sum_k\left[\sigma_k^- - n_k\right]\Gamma_k^{\mathrm{s}}\mathbf{v}   \\\nonumber
&&+\!\!\sum_{k, m\neq k}\!\!\left[\sigma_k^+\sigma_m^- - (1-n_k)n_m \right]  \Gamma_{km}^{\mathrm{d}} \mathbf{v} + \!\!\sum_{k, m\neq k} \!\!  \left[\sigma_k^-\sigma_m^+-n_k(1-n_m) \right]  \Gamma_{km}^{\mathrm{d}} \mathbf{v}\\\nonumber
&&+\!\!\sum_{k, m\neq k}\!\!\left[\sigma_k^-\sigma_m^--n_kn_m \right]  \Gamma_{km}^{\mathrm{d}}  \mathbf{v}+\!\!\sum_{k, m\neq k}\!\!\left[\sigma_k^+\sigma_m^+-(1-n_k)(1-n_m) \right]  \Gamma_{km}^{\mathrm{d}}  \mathbf{v}.
\end{eqnarray}

Here, the single-flip $\Gamma_k^{\mathrm{s}}$ and double-flip rates $\Gamma_{km}^{\mathrm{d}}$ are given by
\begin{eqnarray*}
\Gamma_k^{\mathrm{s}}&=&\Gamma_k+\Omega^4\beta_k -\Omega^4 \suml{m\neq k\atop i=1,4}{} \lt  \sx_k R_i^{km} {R'}_i^{km}\sx_k   + R_i^{mk} {R'}_i^{mk}   \rt\\
   &&-\Omega^4  \suml{m\neq k\atop i=2,3}{} \lt  \sx_k R_i^{mk} \sx_k {R'}_i^{mk}   + R_i^{km} {R'}_i^{km}   \rt     \\
\Gamma_{km}^{\mathrm{d}} &=& \Omega^4\suml{k,m\neq k\atop i=1,4}{}  \sx_k R_i^{km} {R'}_i^{km}\sx_k + \Omega^4\suml{k, m\neq k\atop i=2,3}{}  \sx_m R_i^{km}  \sx_m {R'}_i^{km}  .
\end{eqnarray*}
Conservation of probability is ensured by the preservation of the trace discussed above. The remaining requirement for obtaining a proper classical rate equation is that it must preserve positivity as well, i.e., probabilities cannot become negative. This is equivalent to requiring that every stochastic rate must be positive. Within the perturbative regime, this is automatically satisfied for all single spin-flip processes, since the second-order rates $\Gamma_k$ constitute the leading terms and are strictly positive for all $k$. On the other hand, the two spin-flip ones can become negative for some choices of the parameters, signalling the breakdown of this simplified stochastic interpretation.

\begin{figure}
\includegraphics[width=\columnwidth]{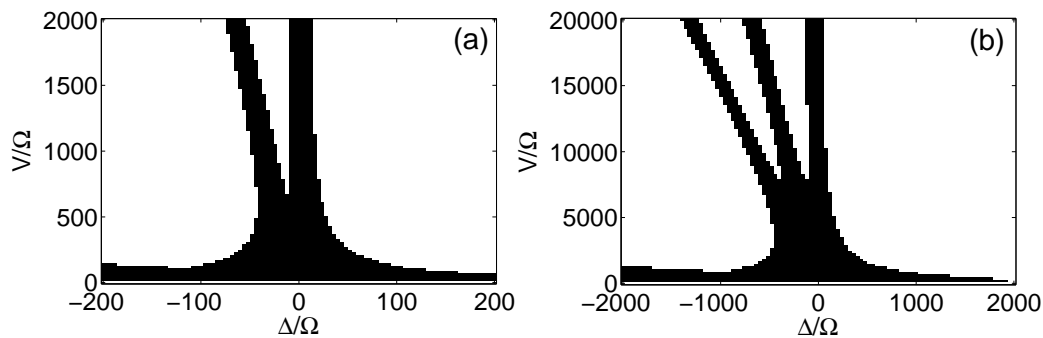}
\caption{$V-\Delta$ regimes in which all rates in \reff{eq:upto4} up to fourth order are positive (white) and where some rates are negative (black) for $\gamma=10\Omega$ and \textbf{(a)}: $N=4$ and \textbf{(b)}: $N=9$.}\label{fig:negative}
\end{figure}

In figure \ref{fig:negative} we show these two regimes emerging from a numerical analysis in the $V$-$\Delta$ plane for $\gamma=10\Omega$ and two system sizes: $N=4$ and $N=9$. The white area corresponds to parameter choices for which all rates are positive, whereas within the black one at least one is negative. We note that typically both $V$ and $\Delta$ must be quite large compared to $\Omega$ in order for the stochastic interpretation to formally hold. We also observe that the boundaries of the ``negative (black) region'' shift towards larger values of both $V$ and $\Delta$ as the system size is increased. In passing, we remark that, as it should be expected, in the non-interacting case $V=0$ the rates for two spin-flip processes vanish, leaving finite only the corrections to the second-order term.

It is worth mentioning that, even when positivity is not ensured for all times, this approach yields a non-negligible reduction of the degrees of freedom. We have numerically integrated the fourth-order equation \reff{eq:upto4} and compared it with the full quantum evolution. The corresponding results for the evolution of the average density, its fluctuations and the density-density correlations are shown in figure \ref{fig:dephasing}. We can see here that, as expected within our perturbative scheme, the fourth-order terms give rise to very small corrections in the dynamics.

\subsection{Perturbative treatment of the radiative decay}

Additional processes can in principle be included in this treatment as long as they do not violate the separation of time-scales. For instance, spontaneous radiative decay from the excited ($\ket{\uparrow}$) to the ground state ($\ket{\downarrow}$) with a rate $\Gamma_\mathrm{ryd}$ can be modelled via a Markovian dissipator
\be
	\mal{D}_\mathrm{dec} \rho = \Gamma_\mathrm{ryd} \suml{k}{}\lt \sigma^-_k \rho \sigma^+_k - \frac{1}{2} \acomm{n_k}{\rho} \rt,
	\label{eq:decay}
\ee
which commutes with the one in \reff{eq:dissipator}, i.e.,
\begin{equation*}
	\mal{D} \mal{D}_\mathrm{dec} \rho = \mal{D}_\mathrm{dec}  \mal{D} \rho.
\end{equation*}
Therefore, the typical time-scale $1/\gamma$ due to dephasing is unaffected and one can still analyse the projected dynamics in the corresponding stationary (diagonal) ensemble. However, decay does not commute with the interaction term and thus introduces an evolution which is not easy to account for analytically. This issue can be overcome by also considering the decay as a slow process or a perturbation, i.e. $\Gamma_\mathrm{ryd} \ll \gamma$. Up to the first non-trivial order in both the decay rate and the coherent flipping amplitude $\Omega$ the effective rate equation reads
\be
	\dot{\mathbf{v}}=\sum_k\Gamma_k\left[\sigma_k^+-(1-n_k) \right]\mathbf{v}+\sum_k\left(\Gamma_k+\Gamma_\mathrm{ryd}\right) \left[\sigma_k^- - n_k\right]\mathbf{v},
	\label{eq:withdec}
\ee
with $\Gamma_k$ as in \reff{eq:Gammak}. Note that $\Gamma_\mathrm{ryd}$ in \reff{eq:decay} must be positive and therefore \reff{eq:withdec} constitutes a proper classical master equation.

\section{Three-level Rydberg atoms in a EIT configuration}\label{sec:EIT}

The second scenario we consider is a frozen gas of $N$ atoms with three internal levels subject to EIT conditions, as shown in figure \ref{fig:schemes}(b). Here the ground state $\ket{\downarrow}$ is resonantly coupled to an intermediate state $\ket{\leftarrow}$ via a laser field with Rabi frequency $\Omega_p$. A second laser couples $\ket{\leftarrow}$ to $\ket{\uparrow}$ with Rabi frequency $\Omega_c$ and detuning $\Delta$. Once again, we only account for interactions $V_{km}$ between pairs of atoms in the Rydberg state. Overall, the Hamiltonian that governs the coherent evolution of this many body system can be expressed as $H=H_0+H_1$, with
\begin{equation*}
	H_0 = \Delta\sum_{k}n_k+   \ha \suml{k\neq m}{}  V_{km} n_k n_m,
\end{equation*}
where $n_k$ denotes the occupation number of the $k$-th Rydberg level, and
\begin{equation*}
	H_1 = \suml{k}{} \lqq  \Omega_p \lt \ket{\downarrow_k} \bra{\leftarrow_k} + \ket{\leftarrow_k} \bra{\downarrow_k}  \rt   +  \Omega_c \lt  \ket{\leftarrow_k} \bra{\uparrow_k} +  \ket{\uparrow_k} \bra{\leftarrow_k}   \rt    \rqq.
\end{equation*}
Atoms excited to a Rydberg level are typically quite stable and display mesoscopic lifetimes of the order of tens of $\mu s$ \cite{Low12}. Thus, on microscopic time-scales the main process causing loss of energy is spontaneous radiative decay of the intermediate state $\ket{\leftarrow}$ to the ground state $\ket{\downarrow}$, which occurs with rate $\Gamma$. We model such a source of dissipation as
\begin{equation*}
	\mal{D} \rho = \Gamma \suml{k}{} \lt \ket{\downarrow_k} \bra{\leftarrow_k} \rho \ket{\leftarrow_k} \bra{\downarrow_k} - \ha \acomm{\proj{\leftarrow_k}}{\rho}   \rt,
\end{equation*}
where we have again neglected spatial and temporal correlations, i.e., each atom decays independently of the state of the others.

\subsection{Second order effective evolution}

We assume now that $\Gamma$ is much larger than both Rabi frequencies $\Omega_c$ and $\Omega_p$. In this limit the population of the intermediate state $\ket{\leftarrow}$ will decay on a fast timescale $\Gamma^{-1}$ and thus one can adiabatically eliminate it. We can then describe the system's state by means of a reduced density matrix $\mu$ which includes only the two internal states $\ket{\downarrow}$ and $\ket{\uparrow}$. Note that in this case coherences between the Rydberg and ground states are preserved. Despite the fact that a classical interpretation is no longer possible, this approach yields a considerable reduction in the growth of the Hilbert space dimension with the system size (from $3^N$ to $2^N$). This can prove useful for numerical approaches focussing on the aforementioned subspace. In this case the observables one is effectively restricted to are those which can be written as combinations of
\begin{eqnarray*}
	n_k = \proj{\uparrow_k} &  \quad , \quad    \sx_k = \ket{\uparrow_k} \bra{\downarrow_k}  + \ket{\downarrow_k} \bra{\uparrow_k} \quad \mathrm{and}  \nol
	& \sy_k = -i\ket{\uparrow_k} \bra{\downarrow_k}  + i\ket{\downarrow_k} \bra{\uparrow_k} .
\end{eqnarray*}
A detailed discussion on how to implement this approximation can be found in \ref{app:EIT}.

First, let us note that $H_0$ is entirely written in terms of the operators $n_k = \proj{\uparrow_k}$, which are not directly affected by the dissipation. Hence, defining $\mal{H}_0\bullet = -i\comm{H_0}{\bullet}$, we find $\comm{\mal{D}}{\mal{H}_0}=0$. However, in contrast with the previous case, the stationary subspace of $\mal{D}$ is not entirely included in the one of $\mal{H}_0$, which implies that $H_0$ generates a non-trivial dynamics within it. Because of this fact, it becomes more difficult to account for its presence (although it is still possible to do it analytically, as we mention at the end of Appendix C). In the following, we shall treat both $H_0$ and $H_1$ perturbatively, which yields, up to second order, the effective equation for the reduced density matrix $\mu$
\be \label{eq:dotmu}
\dot{\mu} = -i\comm{H_0}{\mu} + \suml{k}{} L_k \mu L_k^\dag - \ha \acomm{L_k^\dag L_k}{\mu}.
\ee
This is a Lindblad quantum master equation with coherent part governed by $H_0$ and dissipation provided by the jump operators
\begin{equation*}
	L_k=\frac{2}{\sqrt{\Gamma}}\lt\Omega_c\sigma_-^k+\Omega_p p_k\rt,
\end{equation*}
with $p_k = \proj{\downarrow_k}$ being the projector onto the ground state in the reduced space. Note that here, whilst the coherent part $H_0$ simply acts onto the ``classical'' configurations by associating a given energy to each of them, it is the jump operators that tend to bring the system into coherent superposition states. The engineering of such a type of dissipation, namely one that leads to dynamics and stationary states featuring quantum coherence and many-body superpositions, has been attracting an increasing amount of attention recently \cite{Kraus08,Diehl08,Diehl11,Bardyn12,Schindler13}.

\begin{figure}
\includegraphics[width=0.9\columnwidth]{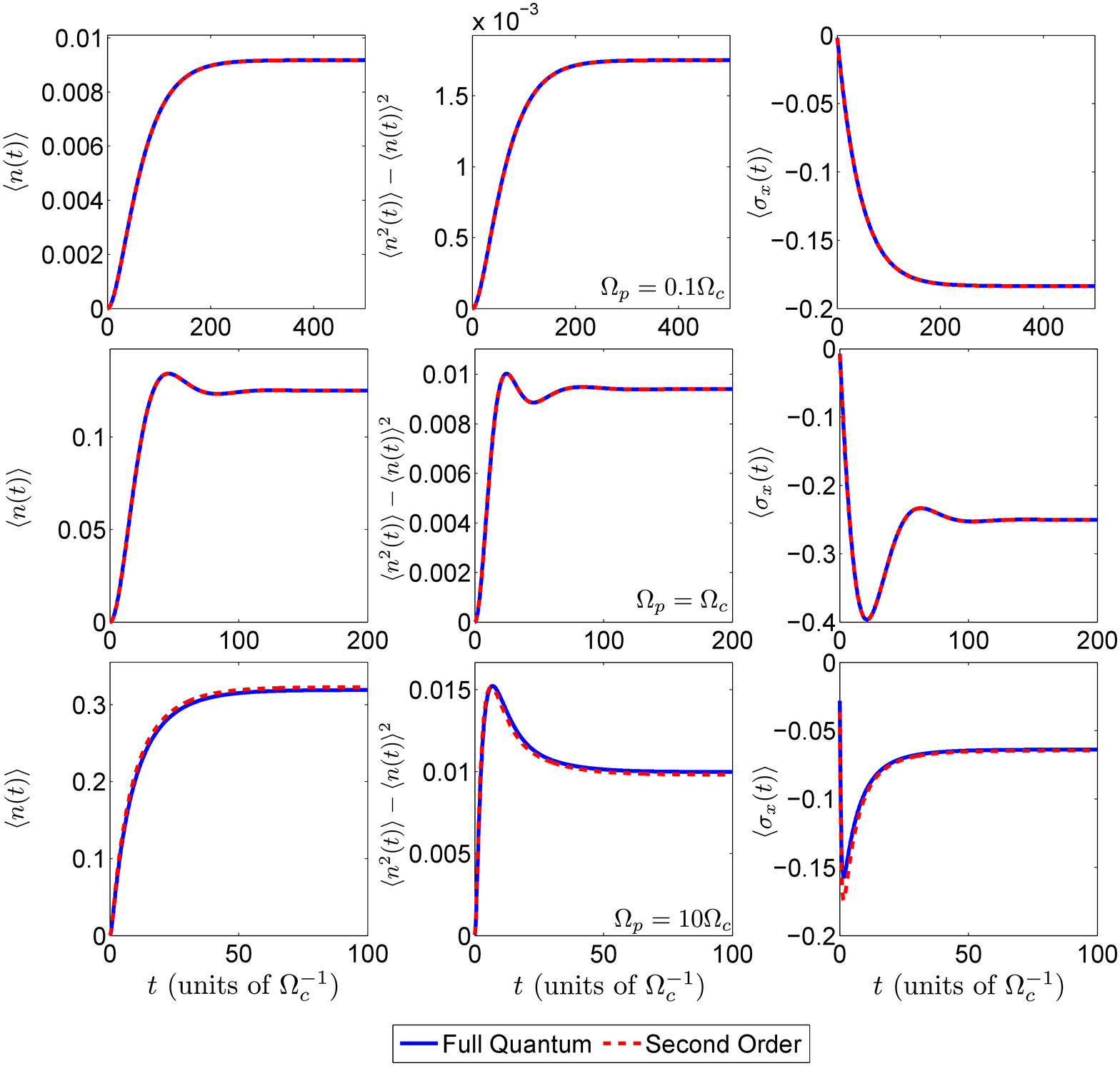}
\caption{Time evolution of the density of excitations $\langle n\rangle$, its associated fluctuations $\langle n^2\rangle-\langle n\rangle^2$ and the coherence, measured here by $\langle\sigma_x\rangle$. In all cases the initial state is the one without excitations $\otimes_k \ket{\downarrow_k}$. We compare the results obtained from the numerically exact integration of the full quantum system and the effective Master equation obtained up to second order [given by \reff{eq:dotmu}] for $N=5$ atoms. The parameters used in the simulations shown are $\Delta=0$, $\Gamma=100\Omega_c$, $V=100\Omega_c$ and $\Omega_p/\Omega_c=0.1, 1, 10$.}
	\label{fig:EIT}
\end{figure}

In order to assess the validity of this effective equation of motion, we resort again to the numerical simulation of a one-dimensional system on a periodic lattice of spacing $a$ and van der Waals interaction, with the nearest-neighbour interaction denoted by $V=C_6/a^6$. We have numerically solved both the full quantum many-body and the second order effective dynamics for small systems of up to $N=6$ atoms. In both cases we have employed a numerically exact direct integration of the corresponding master equations, which quickly becomes very demanding in terms of computational resources as the system size is increased. In figure \ref{fig:EIT} we display the time evolution of the expectation value of the Rydberg density and its fluctuations for the resonant ($\Delta = 0$) case, $\Gamma=V=100\Omega_c$ and three different values of $\Omega_p/\Omega_c=0.1,1,\,\mathrm{and}\,10$. As discussed above, in contrast with the previous two-level system [figure \ref{fig:schemes}(a)] here we keep also track of the dynamics of some off-diagonal observables, such as the operator $\sigma_x=\sum_k \sx_k/N$ (see third column in figure \ref{fig:EIT}). Moreover, we display not only the short-time behaviour but also the long-time one, as in this case the stationary state is not trivial and generically depends on the parameters of the system (in particular on the Rabi frequencies $\Omega_c$ and $\Omega_p$). As a consequence, in this case the steady state will in principle only be reproduced in a perturbative fashion. We observe in general good agreement between the approximate and the exact results. In particular, the curves are hardly distinguishable for $\Omega_p/\Omega_c=0.1$ and $1$. Small deviations show up instead in the case $\Omega_p/\Omega_c=10$, which are related to the fact that $\Gamma$ is only 10 times larger than $\Omega_p$ and therefore we are approaching the limits of applicability of our perturbative scheme. As long as one chooses parameters in a range compatible with the latter, however, we can conclude that \reff{eq:dotmu} provides a good approximation for the description of the dynamics of an interacting Rydberg gas under EIT conditions.

\subsection{Nearest-neighbour exclusion}

A further simplification can be obtained in a particularly simple case: a one-dimensional chain of atoms with resonant excitation ($\Delta=0$) where we approximate the interaction as a hard-wall repulsion between neighbouring excitations, i.e.,
\begin{equation*}
	H_0 = \lim_{V \to +\infty}  \ha \suml{\av{kq}}{} V n_k n_q.
\end{equation*}
This effectively yields a projection of the dynamics onto the set of ground states of $H_0$, i.e., the portion of the Hilbert space spanned by classical states without neighbouring pairs of excitations. This nearest neighbour exclusion approximation has been often used for gaining insight on the underlying physics of strongly-interacting Rydberg gases \cite{Ji11,Ates12-1,Petrosyan13,Hoening13}.

One can show that, for any initial condition with overlap only on allowed (zero-energy) states, the effective equation for the dynamics of the reduced density matrix $\mu$ has a purely dissipative form \cite{Olmos14}, i.e.
\be \label{eq:perfect}
\dot{\mu} = \suml{k}{}J_k \mu J_k^\dag - \ha \acomm{J_k^\dag J_k}{\mu}
\ee
with
\begin{equation*}
J_k=\frac{2}{\sqrt{\Gamma}}\lt\Omega_c\mal{P}_k\sigma_-^k+\Omega_p p_k\rt.
\end{equation*}
Here we have introduced the operator $\mal{P}_k=p_{k-1}p_{k+1}$, which indeed ensures that an excitation $\ket{\uparrow}$ is never created next to an already existing one.

\begin{figure}
\includegraphics[width=\columnwidth]{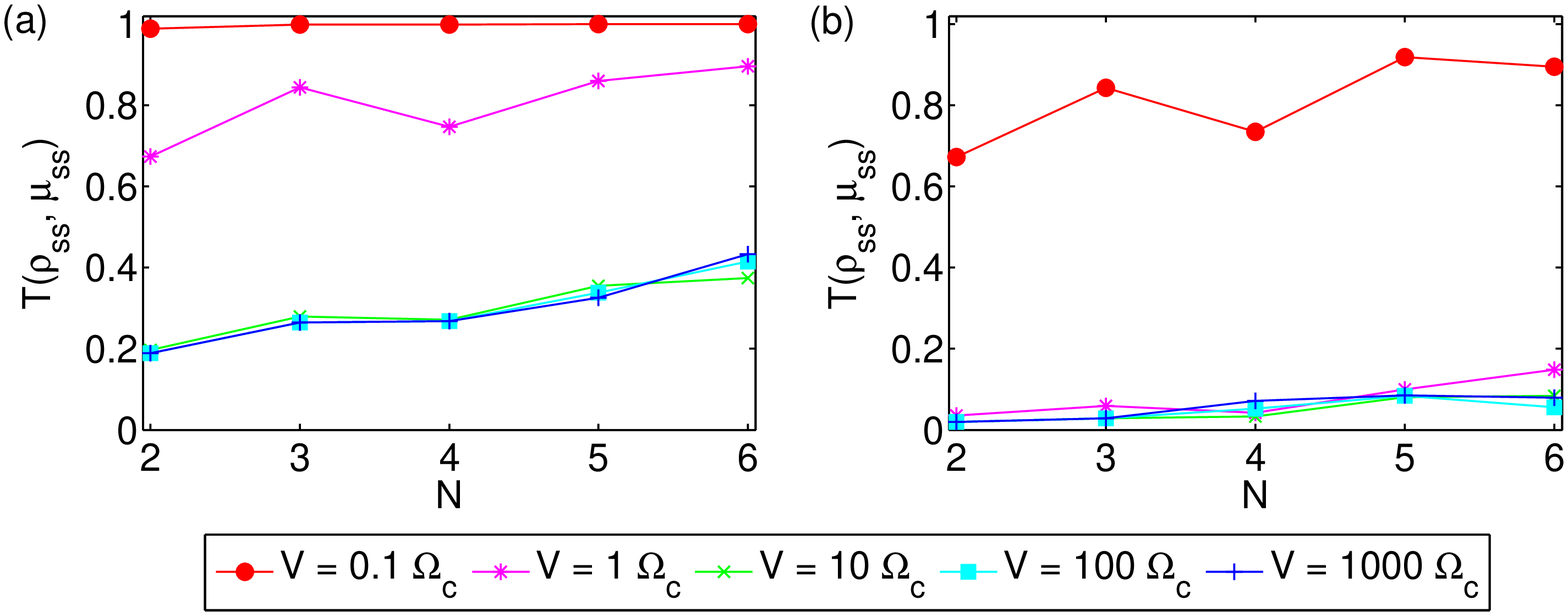}
\caption{The validity of \reff{eq:perfect} as a description of the three-level systems with nearest neighbour exclusion is tested for different parameter regimes and different system sizes up to $N=6$. Curves of different colours represent the trace distance $T(\rho_\mathrm{ss},\,\mu_\mathrm{ss})$ \reff{eq:tdist} between the steady states of the full quantum system and the two-level approximation for different values of the interaction. The two panels show the cases \textbf{(a):} $\Gamma/\Omega_c=100$ and \textbf{(b):} $\Gamma/\Omega_c=1000$. In both panels the value of $\Omega_p$ is $10\Omega_c$.}
\label{fig:size}
\end{figure}

In figure \ref{fig:size} we assess the validity of this approximation by numerical methods. This time we use as a measure the trace distance of two density matrices $\rho$ and $\mu$
\be
	T(\rho,\,\mu) = \ha \Tr{ \sqrt{ (\rho - \mu )^2 } }.
	\label{eq:tdist}
\ee
In particular, we calculate the trace distance between the stationary state of the full three-level many-body system with nearest neighbour interaction $V$, $\rho_\mathrm{ss}$, and the one obtained from the approximate equation \reff{eq:perfect}, $\mu_\mathrm{ss}$. In figure \ref{fig:size} we plot $T(\rho_\mathrm{ss},\,\mu_\mathrm{ss})$ as a function of the system size $N$ for $\Omega_p/\Omega_c=10$ and $\Gamma/\Omega_c=100$ [panel (a)] and $1000$ [panel (b)], as well as for different values of the interaction $V$. First, we observe that the validity of the approximation appears to get slightly worse as the system size is increased. Secondly, while for low values of the interaction $V$ the approximation is poor (as expected), the trace distance rapidly reaches a saturation value when increasing $V$. Finally, by comparing panels (a) and (b) we infer that this saturation value tends to vanish when the expansion parameters $\Omega_{p/c} / \Gamma$ of the perturbative series are made smaller, which is indeed compatible with the fact that we expect the stationary state to be only perturbatively reproduced to second order in our treatment.

\section{Conclusions}

In this paper, we have applied the Nakajima-Zwanzig projection technique to the study of strongly-interacting many-body dynamics, particularly in the context of Rydberg gases. By relying on a time-scale separation between a fast and a slow dynamics, effectively integrating out the fast degrees of freedom and using perturbation theory, we have obtained effective equations of motion that approximately describe the dynamics of a selected set of observables of the system within a reduced subspace. This yields a reduction of the complexity of the corresponding problem, which allows for a numerical treatment of larger systems. Via numerical simulations of small systems we have verified that the obtained effective dynamics yields indeed an excellent approximation to both the stationary state and the relaxation towards it.

We have focussed in particular on two models which describe many-body systems which are currently intensely investigated and experimentally realised with strongly-interacting Rydberg gases. The first one considers each atom as a subsystem with only two physically relevant levels coupled by a laser and subject to strong dephasing. Here, we have found that a classical effective description of the problem is possible, thereby making very large systems amenable to numerical treatment. In the second case we have considered a Rydberg gas under EIT conditions where the fast evolving timescale is provided by the rapid decay of the intermediate level. In this latter case we have also obtained a reduction of the complexity of the problem, whilst not as significant as in the previous case, since part of the Hilbert space structure is retained.

The main aim of this work is to provide a formal framework for the effective description of these strongly interacting systems in the limit of strong dissipation. We hope this effort to contribute towards unifying different results already obtained in the literature (e.g. \cite{Ates06,Ates07-2,Lesanovsky13,Olmos14}) and, moreover, provide some degree of guidance and a reference for future efforts employing these techniques in the context of interacting Rydberg gases.

\ack
The authors would like to thank J.P. Garrahan for some insightful discussions. The research leading to these results has received funding from the European Research Council under the European Union's Seventh Framework Programme (FP/2007-2013) / ERC Grant Agreement No. 335266 (ESCQUMA) and the EU-FET Grant No. 512862 (HAIRS). B.O. acknowledges funding from the University of Nottingham.

\appendix

\section{Adiabatic elimination of fast degrees of freedom}\label{app:general}

In this appendix we establish the general formalism and notation we have employed to derive the reduced dynamical equations in the main text. It is based on the Nakajima-Zwanzig projection formalism \cite{Nakajima58, Zwanzig60}, which relies on finding a criterion to divide the degrees of freedom in relevant and irrelevant, and effectively keeping track only on the former. In a quantum setting, the most intuitive application of this frame of thought would be to focus the attention on a subsystem, while treating the rest as an effective ``external'' bath \cite{Breuer99, Hartmann14}. Here, however, we shall take a slightly different perspective and hinge instead upon a clear time-scale separation for different dynamical processes.

Let us consider a general system whose Markovian dynamics is described by the von Neumann equation
\be
	\dot{\rho} = \lt \LO + \LI \rt \rho,
	\label{eq:evol1}
\ee
where $\LO$ and $\LI$ are two time-independent Liouville operators acting on the density matrix $\rho$. Our aim is to project the evolution onto a reduced subspace by adiabatically eliminating the fast degrees of freedom, which we assume to be entirely described by $\LO$. We also assume for simplicity that $\LO$ includes a dissipative part. In other words, we are trying to obtain a coarse-grained equation of motion which effectively captures the dynamics of the system on time scales longer than the typical ones of $\LO$. Within this frame of thought, we introduce the projector
\begin{equation*}
	P = \liml{T \to +\infty}  \int_{0}^T \frac{\rmd t}{T} \rme^{t \LO}    .
\end{equation*}
onto the stationary subspace of $\LO$ (i.e., its null eigenspace, or kernel). The existence of the limit is ensured by the fact that positivity must be preserved by $\LO$ and therefore all its eigenvalues must have non-positive real part. The reduced density matrix whose dynamics we want to describe is thus $\mu = P \rho$. We correspondingly define the complementary projector $Q = \id - P$, where $\id$ is the identity superoperator, and $\chi = Q\rho$.

Applying $P$ and $Q$ to \reff{eq:evol1}, we can rewrite it as
\be
	\sysb{l}
		\dot{\chi} = Q \lt \LO  + \LI \rt \chi + Q\LI \mu \\[2mm]
		\dot{\mu} = P\LI \mu + P\LI \chi,
	\syse
	\label{eq:projevol1}
\ee
where we have used the fact that, by construction, $P\LO =  \LO P = 0$. An implicit solution of the first equation is given by
\begin{equation*}
	\chi(t) = \rme^{Q \lt \LO  + \LI \rt} \chi (0) + \int_0^t \rmd \tau \,\, \rme^{(t-\tau) Q (\LO + \LI)} Q \LI \mu(\tau).
\end{equation*}
This allows us to write down an integro-differential equation for $\mu$ which does not depend on $\chi(t)$, but only on its initial value. If we further assume that the initial condition entirely lies within the kernel of $\LO$, i.e., $\chi(0) = 0$, we can rewrite the equation for $\mu$ as
\be
	\dot{\mu} = P\LI \mu + P\LI \int_0^t \rmd \tau \,\, \rme^{(t-\tau) Q (\LO + \LI)} Q \LI \mu(\tau).
	\label{eq:proj1}
\ee

The equation above is still exact, but its integro-differential nature makes it difficult to approach analytically. We thereby proceed by applying a Laplace transform ${\bf {\tt L}}[\bullet]=\int_0^\infty \rmd t \rme^{-st} (\bullet)$, as it readily yields a perturbative expansion. This leads to
\be
	s \hat{\mu} (s) - \mu(0) = P\LI \left(\id+\frac{1}{s - Q (\LO + \LI)} Q \LI \right)\hat{\mu}(s),
	\label{eq:Laplace}
\ee
where we have employed some of the general properties of Laplace transforms, i.e.,
\begin{equation*}
	{\bf {\tt L}} [\dot{\mu}] = \hat{\mu}(s) - \mu(t=0),\quad {\bf {\tt L}} [\rme^{tA}] = \frac{1}{s-A},\quad {\bf {\tt L}} [f \ast g] = \hat{f} (s) \hat{g} (s),
\end{equation*}
where $f\ast g$ denotes the convolution $\int_0^t \rmd \tau \, f(\tau) g(t-\tau)$. By assuming that the amplitude of $\LI$ is much smaller than the other energy scales (which makes the corresponding dynamics much slower), we can expand and subsequently truncate the fraction in \reff{eq:Laplace} as a power series in $\LI$:
\begin{eqnarray*}
	\frac{1}{s - Q (\LO + \LI)}&=&\frac{1}{s - Q\LO}\frac{1}{\id - \lt s - Q\LO \rt^{-1} Q\LI} \\
	&=&\frac{1}{s - Q\LO}\suml{j=0}{\infty} \lqq  \lt s - Q\LO \rt^{-1} Q\LI   \rqq^j.
\end{eqnarray*}
Note that, since both $\LO$ and the sum $\LO + \LI$ are assumed to be proper evolution superoperators, their spectra (and thus the singularities of the Laplace transform above) lie to the left of the imaginary axis. It is therefore possible to easily choose a contour on which to define the Laplace anti-transform. By exploiting the same general properties seen above, one can rewrite \reff{eq:proj1} in powers of $\LI$ as
\begin{equation}
   \dot{\mu} =\! \sum_{\alpha=1}^\infty \!\! \lt  {\cal L}^{(\alpha)}\mu  \rt\!\! (t)= P\LI\sum_{\alpha=1}^\infty \prod_{k=1}^{\alpha-1}\lqq\int_0^{\tau_{k-1}}\!\! \rmd \tau_k\,\, \rme^{(\tau_{k-1}-\tau_k)\LO} Q \LI\rqq \mu(\tau_{\alpha-1}),\label{eq:terms1}
\end{equation}
where $\tau_0\equiv t$.

In order to obtain a differential equation (i.e., an expression for $\dot{\mu} (t) $ which only depends on $\mu(t)$), we perform the trivial substitution $\mu(\tau_{\alpha-1})=\mu(t)+\left[\mu(\tau_{\alpha-1})-\mu(t)\right]$ and express the second addend as
\be
	 \mu(\tau_{\alpha-1}) - \mu(t) = \int_{t}^{\tau_{\alpha-1}} \rmd \tau \, \dot{\mu}(\tau)
	=  \sum_{\beta=1}^\infty\int_t^{\tau_{\alpha-1}} \rmd \tau   \lt {\cal L}^{(\beta)}\mu  \rt(\tau).
	\label{eq:diff}
\ee
Note that the difference above is of the same order of the derivative, i.e., at least $O(\LI)$, and that it can be made time-local to any finite perturbative order by iteration: for instance, up to second order it can be cast as
\begin{eqnarray*}
	\mu(\tau_{\alpha-1}) - \mu(t) &\approx&  \int_{t}^{\tau_{\alpha-1}} \rmd \tau P \LI\mu(t)\\
&+&\!\int_{t}^{\tau_{\alpha-1}} \rmd \tau P \LI\Biggl\{ \int_0^\tau \rmd \tau' \, \rme^{(\tau - \tau')\LO} Q \LI  + \int_t^\tau \rmd \tau' \, P \LI\Biggr\} \mu(t).
	\label{eq:diff1}
\end{eqnarray*}
This constitutes substantially an adaptation to the present case of the time convolutionless technique (see e.g., \cite{Breuer99}). In fact, we can iteratively apply \reff{eq:diff} to obtain a time-local representation of the corrections to higher orders in the expansion. This leads to a redefinition of the evolution operators ${\cal L}^{(\alpha)}$; the first four orders read
\begin{eqnarray*}
    {\cal L}^{(1)} \mu &=& P\LI\mu(t)\\
	{\cal L}^{(2)} \mu & = & P\LI \int_0^t \rmd \tau_1\,\, \rme^{\tau_1\LO} Q \LI \mu(t) \\
	{\cal L}^{(3)} \mu & = & P\LI \int_0^t \rmd \tau_1 \,  \rme^{\tau_1\LO} Q \LI  \left[\int_0^{t-\tau_1} \rmd \tau_2 \,  \rme^{\tau_2\LO} Q -\int_{-\tau_1}^0 \rmd \tau_2 \,  P \right] \LI\mu(t)\\
	{\cal L}^{(4)} \mu & =& P\LI  \int_0^t \rmd \tau_1 \,  \rme^{\tau_1\LO} Q \LI\left[\int_0^{t-\tau_1} \rmd \tau_2 \,  \rme^{\tau_2\LO} Q - \int_{-\tau_1}^0 \rmd \tau_2 \,  P \right]\LI\times\\
&&\left[\int_0^{t-\tau_1-\tau_2} \rmd \tau_3 \,  \rme^{\tau_3\LO} Q
-\int_{-\tau_1-\tau_2}^{0} \rmd \tau_3 \, P \right]\LI \mu(t),
\end{eqnarray*}
where we have applied the set of changes of variables $\tau_{k+1} \to \tau_k - \tau_{k+1}$ and $\tau_1 \to t-\tau_1$.

Note that the projectors $Q$ in front of each exponential $\rme^{\tau_k \LO}$ ensure that the latter can only act on states not belonging to its kernel. We furthermore assume that there are no eigenvalues of $\LO$ which are purely imaginary. This implies that $Q$ projects onto eigenspaces corresponding to eigenvalues with strictly negative real parts, such that the action of $\rme^{\tau_k \LO}$ introduces an exponential dampening typically dictated by the eigenvalue $\lambda$ with the largest non-trivial real part. Our assumption of a clear time-scale separation implies that $-\Re (\lambda)$ must be large compared with the typical energy scales of $\LI$ or, more precisely, larger than those associated to operators which couple the stationary subspace of $\LO$ to its complement. In the light of this, extending the integration domains to the whole real axis should introduce only a small correction. After this approximation, the final form of the terms up to fourth order is
\begin{eqnarray*}
    {\cal L}^{(1)}  \mu &=& P\LI\mu(t)\\
	{\cal L}^{(2)}  \mu &\approx & P\LI \int_0^\infty \rmd \tau_1\,\, \rme^{\tau_1\LO} Q \LI \mu(t)\\
	{\cal L}^{(3)}   \mu &\approx & P\LI \int_0^\infty \rmd \tau_1 \,  \rme^{\tau_1\LO} Q \LI  \left[\int_0^\infty \rmd \tau_2 \,  \rme^{\tau_2\LO} Q -\tau_1  P \right] \LI\mu(t)\\
	{\cal L}^{(4)}  \mu &\approx & P\LI  \int_0^\infty \rmd \tau_1 \,  \rme^{\tau_1\LO} Q \LI\left[\int_0^\infty \rmd \tau_2 \,  \rme^{\tau_2\LO} Q - \tau_1  P \right]\LI\times\\
&&\left[\int_0^\infty \rmd \tau_3 \,  \rme^{\tau_3\LO} Q
-\left(\tau_1+\tau_2\right) P \right]\LI \mu(t).
\end{eqnarray*}

Before applying these expressions to the specific systems mentioned in the main text, let us briefly discuss which observables can be calculated within this scheme. The expectation value of a generic observable $\mal{O}$ is given by $\Tr{ \lqq\mal{O} \rho(t)\rqq}$, whereas within the reduced space one can only calculate $\Tr{ \lqq\mal{O}  \mu(t)\rqq} = \Tr{ \lqq\mal{O} P\rho(t) \rqq}$. Clearly, one can extract information on $\mal{O}$ in the reduced scheme if and only if
\begin{equation*}
	\Tr{ \lqq\mal{O} \rho(t)\rqq} = \Tr{ \lqq\mal{O} P\rho(t) \rqq}.
\end{equation*}
Exploiting the fact that the trace defines a scalar product in the superoperatorial space
\begin{equation*}
	\lt  \rho , \sigma \rt \equiv \Tr{\lqq\rho^\dag \sigma\rqq}
\end{equation*}
we can rewrite the relation above as
\begin{equation*}
	\Tr{ \lqq\mal{O} \rho(t)\rqq}= \Tr{\lqq\lt P^\dag \mal{O}  \rt  \rho(t)\rqq},
\end{equation*}
which should be valid for every possible choice of the density matrix $\rho$. Hence, we conclude that only observables that satisfy $\mal{O} = P^\dag \mal{O}$ can be calculated within the reduced-space formalism discussed here.

\section{Two-level Rydberg atoms in the presence of strong dephasing}\label{app:dephasing}

In this Appendix we will give detailed account of the derivation of an effective equation of motion of a system of $N$ driven two-level atoms strongly interacting and in the presence of strong dephasing, as described in Section \ref{sec:dephasing}.

The dynamics of the system is modelled via the master equation \reff{eq:evol1}, where
\be
	\sysb{l}
		\LO \rho = -i\comm{H_0}{\rho} + \mal{D} \rho \\[2mm]
		\LI \rho = \suml{k}{} \LI^k \rho = -i \Omega \suml{k}{} \comm{\sigma_k^x}{\rho}
	\syse
	\label{eq:evoldef}
\ee
with $H_0 = \Delta \suml{k}{} n_k + \frac{1}{2}\suml{k \neq m}{} V_{km} n_k n_m$ and the dissipator $\mal{D} \rho = \deph \suml{k}{}\lt n_k \rho n_k - \frac{1}{2} \acomm{n_k}{\rho} \rt$. The stationary space of $\LO$ is formed here by all matrices $\mu$ which are diagonal in the basis of all possible classical spin configurations. In the following, we will always use the terms ``diagonal'' and ``off-diagonal'' referring to this basis. Strictly speaking, the fast dynamics is provided by the dephasing and thus one would have to define $\LO = \mal{D}$ in order to directly connect to the results of Appendix A. However, in this case the commutator $-i\comm{H_0}{\bullet}$ not only commutes with the dephasing dissipator, but actually its stationary subspace includes the one of $\mal{D}$. Therefore, this term can be conveniently included in $\LO$ without altering the structure of said subspace. On a different note, the perturbation $\LI$ does not connect states belonging to the kernel of $\LO$, i.e., it cannot map any state which is stationary under the action of $\LO$ into another one. This implies that $P \LI P = 0$, and hence the first four terms of the expansion are notably simplified to
\begin{eqnarray*}
    {\cal L}^{(1)}\mu &=& 0\\
	{\cal L}^{(2)}\mu &\approx & P\LI \int_0^\infty \rmd \tau_1\,\, \rme^{\tau_1\LO} Q \LI \mu(t) \\
	{\cal L}^{(3)}\mu &\approx & P\LI \int_0^\infty \rmd \tau_1 \,  \rme^{\tau_1\LO} Q \LI \int_0^\infty \rmd \tau_2 \,  \rme^{\tau_2\LO} Q\LI\mu(t)\\
	{\cal L}^{(4)}\mu &\approx & P\LI  \int_0^\infty \rmd \tau_1 \,  \rme^{\tau_1\LO} Q \LI\left[\int_0^\infty \rmd \tau_2 \,  \rme^{\tau_2\LO} Q - \tau_1  P \right]\LI\times\\
&&\int_0^\infty \rmd \tau_3 \,  \rme^{\tau_3\LO} Q\LI \mu(t).
\end{eqnarray*}

Let us first calculate the second order contribution to the dynamics. To do so, and in order to be able to compute any term in the perturbative expansion, it is fundamental to understand how the operator
\begin{equation*}
	\mal{O} = \int_0^\infty \rmd \tau\,\, \rme^{\tau\LO} Q \LI
\end{equation*}
acts on a diagonal matrix. Note that, according to \reff{eq:evoldef}, one can actually reduce this problem to studying the action of the single-site components
\begin{equation*}
	\mal{O}_k = \int_0^\infty \rmd \tau\,\, \rme^{\tau\LO} Q \LI^k \comma \mal{O} = \suml{k}{} \mal{O}_k.
\end{equation*}
We start by decomposing the diagonal matrix $\mu$ as
\begin{eqnarray}\nonumber
	\mu &=& \mu^k_\uparrow \otimes \proj{\uparrow_k} + \mu^k_\downarrow \otimes \proj{\downarrow_k}\\
&=& \lt \frac{\mu^k_\uparrow + \mu^k_\downarrow}{2} \rt \otimes \id_k + \lt \frac{\mu^k_\uparrow - \mu^k_\downarrow}{2} \rt \otimes \sz_k,
	\label{eq:decomp}
\end{eqnarray}
where
\begin{equation*}
	\mu^k_\uparrow = \bra{\uparrow_k} \mu \ket{\uparrow_k}\comma\mu^k_\downarrow = \bra{\downarrow_k} \mu \ket{\downarrow_k}
\end{equation*}
are $2^{N-1} \times 2^{N-1}$ matrices acting on all sites but the $k$-th one. This representation allows us to more easily calculate the action of $\mal{L}_1^k$ on $\mu$, which reads
\begin{equation*}
	\mal{L}_1^k \mu = -i\Omega \lt \frac{\mu^k_\uparrow - \mu^k_\downarrow}{2} \rt \otimes \comm{\sx_k}{\sz_k} = \Omega \lt \mu^k_\downarrow - \mu^k_\uparrow \rt \otimes \sy_k.
\end{equation*}
As this matrix is entirely off-diagonal, the operator $Q$ effectively acts as the identity when applied to it.

We then have now to compute the action of the superoperator $\rme^{\tau \mal{L}_0}$ on a generic matrix of the form $\mu_{\iota_1, \ldots \iota_m}^{k_1, \ldots k_m} \otimes \sigma_{k_1}^y \otimes \ldots \otimes \sigma_{k_m}^y$, with $\iota_n = \uparrow, \downarrow$ and $k_n = 1 \ldots N$. To this end, we first notice that the action of the Hamiltonian $H_0$ and the dissipator $\mal{D}$ commute, which allows us to factorize the exponential as
\begin{equation*}
	\rme^{\tau \mal{L}_0 } (\bullet) = \rme^{-iH_0 \tau} \lqq \rme^{\tau \mal{D}} (\bullet)  \rqq \rme^{iH_0 \tau}.
\end{equation*}
Let us first analyse the effect of the dissipator: Since the dissipation mechanism acts independently on each site, we can further factorize its action as
\begin{equation*}
	\rme^{\tau \mal{D}} = \prodl{k = 1}{N} \rme^{\tau \gamma \mal{D}_k}
\end{equation*}
with $\mal{D}_k (\bullet) = n_k (\bullet) n_k -(1/2) \acomm{n_k}{(\bullet)}$. One can easily show that the action of the dissipator on the diagonal and off-diagonal components yields
\begin{equation*}
	\mal{D}_k \sz_k = \mal{D}_k \id_k = 0 \quad\mathrm{and}\quad  \mal{D}_k \sigma_k^{x/y} = -\frac{1}{2} \sigma_k^{x/y},
\end{equation*}
respectively. Therefore, one obtains
\be
	\rme^{\tau \mal{D}}  \mu_{\iota_1, \ldots \iota_m}^{k_1, \ldots k_m} \otimes \sigma_{k_1}^y \otimes \ldots \otimes \sigma_{k_m}^y = \rme^{-\frac{\deph \tau m}{2}}  \mu_{\iota_1, \ldots \iota_m}^{k_1, \ldots k_m} \otimes \sigma_{k_1}^y \otimes \ldots \otimes \sigma_{k_m}^y,
	\label{eq:damping1}
\ee
which amounts simply to the multiplication by a damping factor. The action of the coherent part is more involved. In order to give an expression for it as well, we divide the Hamiltonian $H_0$ into two operators: one which does not depend on the indices $k_1 \ldots k_m$ and therefore inconsequentially commutes with the density matrix above, and a part which instead depends on them, i.e.,
\begin{equation}
	h_{k_1, \ldots, k_m} \lt n_{k_1}, \ldots, n_{k_m} \rt = H_0 - H_0\eval{n_{k_1}=0, \ldots, n_{k_m}=0}
	\label{eq:hkdef}
\end{equation}
with
\begin{equation*}
	H_0 \eval{n_k = 0} \equiv   \bra{n_k = 0} H_0 \ket{n_k=0} \otimes \id_k .
\end{equation*}
The action of the Hamiltonian on a matrix of the form \reff{eq:damping1} can be obtained by the realisation that
\begin{equation*}
	h_{k_1, \ldots k_m} \lt n_{k_1}, \ldots, n_{k_m} \rt \sigma_{k_j}^+ = h_{k_1, \ldots, k_m} \lt n_{k_1}, \ldots, n_{k_j} = 1, \ldots n_{k_m} \rt \sigma_{k_j}^+
\end{equation*}
and
\begin{equation*}
	\sigma_{k_j}^+   h_{k_1, \ldots, k_m} \lt n_{k_1}, \ldots, n_{k_m} \rt  = \sigma_{k_j}^+   h_{k_1, \ldots, k_m} \lt n_{k_1}, \ldots, n_{k_j} = 0, \ldots, n_{k_m} \rt,
\end{equation*}
with $\sigma_{k}^{+}=\ket{\uparrow_k}\bra{\downarrow_k}$. We emphasize here that $h_{k_1, \ldots, k_m} (n_{k_j}=1)$ stands for $\bra{n_{k_j}=1}h_{k_1, \ldots, k_m} \ket{n_{k_j}=1} $ and consists of a reduced matrix which does not act on the $k_j$-th subspace. Note that this implies that it commutes with every local operator acting only on it, e.g., $\sx_{k_j}$.

Introducing the function $\mal{N}_p$ with $p=\pm$ such that $\mal{N}_+=1$ and $\mal{N}_-=0$ we find
\begin{eqnarray}
	\rme^{-i \tau H_0} && \, \left\{ \mu_{\iota_1, \ldots \iota_m}^{k_1, \ldots k_m} \otimes \sigma_{k_1}^{p_1} \otimes \ldots \otimes \sigma_{k_m}^{p_m} \right\} \,  \rme^{i \tau H_0}  = \nol
	 &&\rme^{- i \tau \lqq h_{k_1, \ldots, k_m} (\mal{N}_{p_1}, \ldots, \mal{N}_{p_m} )  - h_{k_1, \ldots, k_m} (1-\mal{N}_{p_1}, \ldots, 1-\mal{N}_{p_m} ) \rqq} \,\, \mu_{\iota_1, \ldots \iota_m}^{k_1, \ldots k_m} \otimes \nol
	 && \otimes \sigma_{k_1}^{p_1} \otimes \ldots \otimes \sigma_{k_m}^{p_m},
	 \label{eq:Hformula}
\end{eqnarray}
so that the action of the operator $O_k(\tau)$ [with $\mal{O}_k = \int_0^\infty \rmd \tau O_k(\tau)$] on $\mu$, which will be used for the calculation of higher orders as well, yields
\be
    O_k(\tau)\mu=i\Omega \rme^{-\frac{\deph \tau}{2}}\! \lt\mu^k_\downarrow - \mu^k_\uparrow \rt  \otimes\left\{ \rme^{-i \tau \lqq h_k (0) - h_k (1)  \rqq}  \sigma_k^- \! - \!  \rme^{-i \tau \lqq h_k (1) - h_k (0)  \rqq} \sigma_k^+           \right\}\!.
    \label{eq:Ok}
\ee

With these expressions we can already calculate the contribution to the second order. To do so, first we realize that the action of $O_k(\tau)$ on the diagonal matrix $\mu$ yields an off-diagonal form. As a consequence, due to the presence of a projector $P$ as a last step, one needs the subsequent action of $\mal{L}_1^k \lt\sigma_k^{\pm}\rt = \pm i\Omega\sz_k$ --- i.e., specifically of the $k$-th component of $\LI$ --- in order to recover a diagonal matrix and get a non-vanishing outcome. Thus, the second order contribution yields
\begin{eqnarray*}
	{\cal L}^{(2)}\mu&=&P\LI \mal{O}\mu= P\sum_k\int_0^\infty \rmd \tau \LI^kO_k(\tau) \mu\\
& =&  2\Omega^2 P\sum_k\int_0^\infty \rmd \tau \rme^{-\frac{\deph \tau}{2}} \cos{\left[\tau \left( h_k (0) - h_k (1)  \right)\right]}\lt \mu^k_\downarrow - \mu^k_\uparrow \rt  \otimes \sigma_k^z,
\end{eqnarray*}
which after the integration over time reads
\begin{equation*}
{\cal L}^{(2)}\mu= \sum_k \frac{\Omega^2\deph}{\lt\frac{\deph}{2}\rt^2+\lt\Delta+\sum_{q\neq k}V_{kq}n_q\rt^2}\left(\sigma_k^x\mu\sigma_k^x-\mu\right),
\end{equation*}
as shown in \reff{eq:M2}, where we have used that $h_k(0)=0$ and $h_k(1)=\Delta+\sum_{q\neq k}V_{kq}n_q$.

Let us now look into the calculation of the next orders. Here, we again note that the action of the operator $\mal{O}$ on the diagonal matrix $\mu$ yields an off-diagonal one. The action of $\LO$ does not modify the matrix structure, and thus only the subsequent action of $\LI$ can recover a diagonal matrix that is not annihilated by the final application of the projector $P$. This in turn means that any odd number of $\mal{L}_1$s applied to $\mu$ will always render an off-diagonal matrix, yielding
\begin{equation*}
	\mal{L}^{(2j+1)}\mu \equiv  0 \quad \forall \, j \in \N,
\end{equation*}
and, in particular, $\mal{L}^{(3)}\mu=0$. Moreover, as seen above, for every occurrence of e.g., $\LI^k$, a second $k$-th component must be present, since no other $\LI^q$ with $q \neq k$ acts on the $k$-th subspace and is able to recover the diagonal structure. Hence, the different components of $\LI$ always appear in pairs.

We look now into the calculation of the fourth order contribution to the perturbative expansion. We first split it into two terms as
\begin{equation*}
	{\cal L}^{(4)}\mu =  P\LI\mal{O}^3\mu(t)-P\LI\int_0^\infty \rmd \tau_2 \,  \tau_2\rme^{\tau_2\LO} Q \LI P\LI\mal{O}\mu(t)=\lt A_4 +B_4\rt\mu.
\end{equation*}
Let us focus on the first term, $A_4\mu$: from the discussion above, we know that the only non-zero terms are the ones where the $\LI$ operators come in pairs with equal indices, i.e.,
\begin{eqnarray}
	A_4\mu = & P \suml{k,m \neq k}{} \mal{L}_1^k \lqq \mal{O}_k \mal{O}_m \mal{O}_m +  \mal{O}_m \mal{O}_k \mal{O}_m + \mal{O}_m \mal{O}_m \mal{O}_k  \rqq \mu + \nol
	& + P \suml{k}{} \mal{L}_1^k \mal{O}_k \mal{O}_k \mal{O}_k \mu
	\label{eq:A4}
\end{eqnarray}
On the other hand, the off-diagonal projectors $Q$ included in each $\mal{O}$ (note that $\mal{O} = Q \mal{O}$) prevent the matrix structure from being diagonal at any intermediate step before the last one, which means that any subsequence of $\mal{O}$s which appears on the right (i.e., directly acts on $\mu$), is strictly shorter than the full sequence, and in which all indices of the $\LI$ components can be paired up with each other identically vanishes. Thus, \reff{eq:A4} can be simplified to
\be
	A_4\mu = P \suml{k,m \neq k}{} \mal{L}_1^k \lqq  \mal{O}_m \mal{O}_k \mal{O}_m + \mal{O}_m \mal{O}_m \mal{O}_k  \rqq \mu.
	\label{eq:A4_1}
\ee

We calculate now step by step the action of $O_m(\tau_2)O_k(\tau_1)\mu$, common to both terms in \reff{eq:A4_1}. The first step, $O_k (\tau_1) \mu$, we already calculated for the second order in \reff{eq:Ok}. We have now to apply $O_m (\tau_2) = \rme^{\tau_2 \LO} Q \mal{L}_1^m$. The action of $\mal{L}_1^m$ here does not involve solely the difference $\mu^k_{-} \equiv \mu^k_\downarrow - \mu^k_\uparrow$, but also each operator-valued prefactor. In order to calculate its action we employ the same decomposition used in \reff{eq:decomp} and rewrite $\mu^k_{-}$ as
\begin{equation*}
	\mu^k_{-} = \lt \frac{\mu^{k,m}_{-,\downarrow} + \mu^{k,m}_{-,\uparrow}}{2}  \rt \otimes \id_m -    \lt \frac{\mu^{k,m}_{-,\downarrow} - \mu^{k,m}_{-,\uparrow}}{2} \rt \otimes \sz_m.
\end{equation*}
We now make use of the general identity
\be
	\LI^k (M\otimes \sz_k) = -2 \Omega M \otimes \sy_k = 2i\Omega M \otimes (\sigma_k^+ - \sigma_k^-) \label{eq:lik1}
\ee
which, recalling that we are defining $M^m_- = M^m_\downarrow - M^m_\uparrow$, yields
\begin{equation*}
	\LI^m \lt  \rme^{\pm i \tau_1 h_k(1)} \mu^k_{-} \rt = -i \Omega \lt  \rme^{\pm i \tau_1 h_k(1)} \mu^k_{-} \rt^m_- \otimes \lt \sigma_m^- - \sigma_m^+ \rt.
\end{equation*}
Consequently, we find
\begin{eqnarray*}
	\mal{L}_1^m O_k (\tau_1) \mu  & =& -\Omega^2\rme^{-\frac{\deph \tau_1}{2}}   \left\{    \lt  \rme^{ i \tau_1 h_k(1)} \mu^k_{-} \rt^m_-  \otimes \lt \sigma_m^- - \sigma_m^+  \rt \otimes \sigma_k^-   \right.   \\
	&& - \left.   \lt  \rme^{- i \tau_1 h_k(1)} \mu^k_{-} \rt^m_-  \otimes \lt \sigma_m^- - \sigma_m^+  \rt \otimes \sigma_k^+            \right\}.
\end{eqnarray*}
Hence, the overall effect of $O_m(\tau_2)$ reads
\begin{eqnarray}\nonumber
	O_m(\tau_2) O_k(\tau_1) \mu& = &-  \Omega^2 \rme^{-\frac{\deph \tau_1}{2} - \deph \tau_2}    \left\{ \rme^{i\tau_2  h_{k,m}(1,1)}  \lt  \rme^{ i \tau_1 h_k(1)} \mu^k_{-} \rt^m_-    \otimes \sigma_m^- \otimes \sigma_k^-  \right. \\\nonumber
	&&- \rme^{-i\tau_2 \lt h_{k,m} (0,1) - h_{k,m}(1,0) \rt}  \lt  \rme^{ i \tau_1 h_k(1)} \mu^k_{-} \rt^m_-    \otimes \sigma_m^+ \otimes \sigma_k^- \\\nonumber
	&&- \rme^{-i\tau_2 \lt h_{k,m} (1,0) - h_{k,m}(0,1) \rt}   \lt  \rme^{ - i \tau_1 h_k(1)} \mu^k_{-} \rt^m_-    \otimes \sigma_m^- \otimes \sigma_k^+\\
	&&+\left. \rme^{-i\tau_2 h_{k,m} (1,1)}  \lt  \rme^{ -i \tau_1 h_k(1)} \mu^k_{-} \rt^m_-   \otimes \sigma_m^+ \otimes \sigma_k^+	\right\},
\label{eq:sameproc}
\end{eqnarray}
where we have applied \reff{eq:Hformula} to each addend.

Let us now calculate the first term of the sum in \reff{eq:A4_1}, which means that the next step involves the action of a superoperator $O_k (\tau_3)$. Following the same procedure outlined above, applying the identities
\begin{equation*}
	\LI^k (M\otimes \sigma^\pm_k) = \pm  i \Omega M \otimes \sz_k
\end{equation*}
and introducing the shorthand notation
\be
	Z_k \lqq \mu \rqq \equiv \mu^k_- \otimes \sz_k =\matb{cc}
								\mu^k_\downarrow - \mu^k_\uparrow & 0 \\
								0 & \mu^k_\uparrow - \mu^k_\downarrow
							\mate = \sx_k \mu \sx_k - \mu,
							\label{eq:diagrec}
\ee
one obtains after some algebraic manipulation
\begin{eqnarray*}
	O_k (\tau_3) O_m(\tau_2) O_k(\tau_1) && \mu = -  i \Omega^3\rme^{-\frac{\deph \tau_1}{2} - \deph \tau_2 -\frac{\deph \tau_3}{2}} \\
	&&\times\!   \left\{ - \rme^{i\tau_3 h_m(1)}  \lqq  \rme^{ i\tau_2 h_{k,m}(1,1)} \lt \rme^{i\tau_1 h_k(1)}  Z_k[\mu]    \rt^m_-	  \right. \right. \\
	&&+\!\! \left. \left.  \rme^{-i\tau_2 \lqq h_{k,m} (1,0) - h_{k,m}(0,1) \rqq} \! \lt \rme^{-i\tau_1 h_k(1)}  Z_k[\mu]    \rt^m_-	    \rqq     \otimes \sigma_m^-  \right. \\
	&&\left.  +  \rme^{-i\tau_3 h_m(1)}\!\!\lqq  \rme^{-i\tau_2 \lqq h_{k,m} (0,1) - h_{k,m}(1,0) \rqq} \!\!\lt \rme^{i\tau_1 h_k(1)}  Z_k[\mu]    \rt^m_-	 \right. \right. \\
	&&\left. \left. + \rme^{-i\tau_2 h_{k,m} (1,1)}   \lt \rme^{-i\tau_1 h_k(1)}  Z_k[\mu]    \rt^m_-	      \rqq   \otimes \sigma_m^+ 	 \right\}.
\end{eqnarray*}
As a final step for the calculation of this first term, we have to apply $\LI^m$, which renders
\begin{eqnarray*}
	&\LI^m  O_k (\tau_3)  O_m(\tau_2) O_k(\tau_1) \mu = \Omega^4\rme^{-\frac{\deph \tau_1}{2} - \deph \tau_2 -\frac{\deph \tau_3}{2}}  \\
	&\times\!\left\{  \lqq  \rme^{i\lqq\tau_3 h_m(1)+\tau_2  h_{k,m}(1,1)\rqq} + \rme^{-i\left\{\tau_3 h_m(1)+\tau_2 \lqq h_{k,m} (0,1) - h_{k,m}(1,0) \rqq\right\}} \rqq\!\!Z_m\!\! \lqq \rme^{i\tau_1 h_k(1)}  Z_k[\mu]    \rqq  \right.  \\
	&+\!\! \left. \lqq  \rme^{i\left\{\tau_3 h_m(1)+\tau_2 \lqq h_{k,m} (0,1) - h_{k,m}(1,0) \rqq\right\}}   +  \rme^{ -i\lqq\tau_3 h_m(1)+\tau_2  h_{k,m}(1,1)\rqq }   \rqq \!\!   Z_m\!\! \lqq   \rme^{-i\tau_1 h_k(1)}  Z_k[\mu]    	 \rqq    \right\}.
\end{eqnarray*}

In order to calculate the second contribution to $A_4$ in (\ref{eq:A4_1}), we need to go back to \reff{eq:sameproc} and apply $O_m(\tau_3)$ to it, thus obtaining
\begin{eqnarray*}
	&O_m(\tau_3) O_m(\tau_2) O_k(\tau_1) \mu = -i\Omega^3 \rme^{-\frac{\deph \tau_1}{2} - \deph \tau_2 - \frac{\deph \tau_3}{2}} \\
	& \times\!\! \left\{ - \rme^{i\tau_3 h_k(1)} \!\lqq \rme^{i\tau_2 h_{k,m}(1,1)} +  \rme^{-i\tau_2 \lqq h_{k,m} (0,1) - h_{k,m}(1,0) \rqq}\rqq\!    \lt  \rme^{ i \tau_1 h_k(1)} \mu^k_{-} \rt^m_-   \otimes \sigma_m^z \otimes \sigma_k^- \right. \\
	& +\! \rme^{-i\tau_3 h_k(1)} \left.    \!\lqq \rme^{-i\tau_2 h_{k,m}(1,1)} +  \rme^{i\tau_2 \lqq h_{k,m} (0,1) - h_{k,m}(1,0) \rqq}\rqq \!   \lt  \rme^{ - i \tau_1 h_k(1)} \mu^k_{-} \rt^m_-   \otimes \sigma_m^z \otimes \sigma_k^+  \right\}.
\end{eqnarray*}
Finally, we apply the operator $\LI^k$ and arrive at
\begin{eqnarray*}
	&\LI^k O_m(\tau_3) O_m(\tau_2) O_k(\tau_1) \mu = \Omega^4 \rme^{-\frac{\deph \tau_1}{2} - \deph \tau_2 - \frac{\deph \tau_3}{2}}  \\
	& \times\!\! \left\{  \rme^{i\tau_3 h_k(1)} \lqq \rme^{i\tau_2 h_{k,m}(1,1)} +  \rme^{-i\tau_2 \lqq h_{k,m} (0,1) - h_{k,m}(1,0) \rqq}           \rqq     Z_m \lqq   \rme^{i\tau_1 h_k(1)}  Z_k[\mu]    	 \rqq  \right. \\
	& +\! \rme^{-i\tau_3 h_k(1)} \left.    \lqq \rme^{-i\tau_2 h_{k,m}(1,1)} +  \rme^{i\tau_2 \lqq h_{k,m} (0,1) - h_{k,m}(1,0) \rqq}       \rqq     Z_m \lqq   \rme^{-i\tau_1 h_k(1)}  Z_k[\mu]    	 \rqq  \right\},
\end{eqnarray*}
where we have used the definition \reff{eq:diagrec}.

Note that the remaining time integrations involve only diagonal matrices (in the sense defined at the beginning, i.e., in the classical basis) and can be therefore straightforwardly evaluated. We thus obtain
\be
	A_4\mu =\Omega^4\sum_{i=1}^3\suml{m,k\neq m}{} R_i^{km}Z_m \lqq {R'}_i^{km} Z_k \lqq  \mu  \rqq  \rqq,
\ee
where the first summand comes from the first term in \reff{eq:A4_1} and the other two from the second one. In this expression the operator-valued rates ${R}_i^{km}$ and ${R'}_i^{km}$ read
\begin{equation*}
\sysb{l}
    {R}_1^{km} =\id \\[2mm]
	{R'}_1^{km} = 2\Re \lqq\lt {\Gamma^m_1}^\dagger {\Gamma_2^{km}}^\dagger + \Gamma^m_1 \Gamma_{3}^{km}  \rt {\Gamma_1^k}^\dagger\rqq\\[2mm]
    {R}_2^{km} =\Re\lt\Gamma_1^k\rt\\[2mm]
	{R'}_2^{km} =2\Re\lqq{\Gamma_1^k}^\dagger \lt {\Gamma_2^{km}}^\dagger +\Gamma_3^{km} \rt\rqq\\[2mm]
    {R}_3^{km} =-\Im\lt\Gamma_1^k\rt\\[2mm]
	{R'}_3^{km} =-2\Im\lqq{\Gamma_1^k}^\dagger \lt {\Gamma_2^{km}}^\dagger +\Gamma_3^{km} \rt\rqq
\syse
\end{equation*}
with
\begin{equation}
	\sysb{l}
		\Gamma_1^k = \lqq \frac{\deph}{2} \id + i h_k(1)   \rqq^{-1}\\[2mm]
		\Gamma_{2}^{km} = \lqq \deph \id + i h_{k,m}(1,1)   \rqq^{-1}\\[2mm]
		\Gamma_{3}^{km} = \left\{ \deph \id + i \lqq h_{k,m}(0,1) - h_{k,m}(1,0) \rqq \right\}^{-1},
	\syse
	\label{eq:defGamma}
\end{equation}
as already shown in $\reff{eq:M4}$ in Section \ref{sec:dephasing} of the paper. Note that, as stated above, the indices of the operators ``$h$'' in these expressions denote the subspaces on which they do not act and, as a consequence, $\comm{h_k(1)}{\sx_k} = 0$ and $\comm{h_{k,m}(n_k,n_m)}{\sx_{k/m}} = 0$. This property then trivially transmits to the corresponding $\Gamma$-s. Therefore, as reported in the main text, $R_{2/3}^{km}$ and ${R'}_{2/3}^{km}$ commute with $\sx_k$.

The calculation of the second term of \reff{eq:A4_1} follows analogous steps; the final expression reads
\begin{eqnarray*}
B_4\mu&=&-P\LI\int_0^\infty \rmd \tau_1 \,  \tau_1\rme^{\tau_1\LO} Q \LI P\LI\mal{O}\mu\\
&=& -\suml{k,m}{}  \int_0^\infty\! \!\int_0^\infty \! \! \rmd \tau_1\rmd \tau_2 \,\tau_2 \, P\LI^m \, O_m(\tau_2)  P \LI^k   O_k(\tau_1) \mu.
\end{eqnarray*}
Note that, from the second order calculation, we already know the action of $P \LI^k   O_k(\tau_1)$ on $\mu$, which is indeed diagonal. For the next step (the action of $\LI^m \, O_m(\tau_2)$), we distinguish the cases $m=k$ and $m\neq k$. Let us now first consider the case $m = k$, which yields
\begin{equation*}
	 \LI^kO_k(\tau_2) \LI^k   O_k(\tau_1) \mu=- 8\Omega^4   \rme^{ -\frac{\deph (\tau_1 +\tau_2)}{2} }\! \cosa{ \tau_1 h_k(1)} \cosa{ \tau_2 h_k(1)}   Z_k\left[\mu\right].
\end{equation*}
On the other hand, when $m \neq k$ we obtain
\begin{equation*}
	 \LI^mO_m(\tau_2) \LI^k   O_k(\tau_1) \mu=4\Omega^4\rme^{ -\frac{ \deph (\tau_1 + \tau_2)}{2} }   \!\cosa{ \tau_2 h_m(1)}\!  Z_m \lqq \cosa{ \tau_1 h_k(1)}\! Z_k[\mu]   \rqq
\end{equation*}
Finally, integrating over time the last two expressions we obtain the contribution for $B_4\mu$ to fourth order
\begin{equation*}
	B_4\mu =\sum_k\beta_k Z_k\lqq\mu\rqq+\Omega^4\suml{m,k\neq m}{} R_4^{km}Z_m \lqq {R'}_4^{km} Z_k \lqq  \mu  \rqq  \rqq,
\end{equation*}
with
\begin{equation*}
\sysb{l}
    \beta_{k} =64\Omega^4\,\,\frac{\deph \lqq \deph^2 - 4 h_k(1)^2 \rqq}{\lqq \deph^2 + 4 h_k(1)^2 \rqq^3}\\[2mm]
	{R}_4^{km} = - 32\frac{\deph^2 - 4 h_m(1)^2 }{\lqq \deph^2 + 4 h_m(1)^2 \rqq^2}\\[2mm]
    {R'}_4^{km} =\frac{\deph}{\deph^2 + 4h_k(1)^2},
\syse
\end{equation*}
which indeed coincide with the expressions shown in \reff{eq:M4} in Section \ref{sec:dephasing} of the paper, as one can check via the definitions \reff{eq:defGamma}.

\subsection{Radiative decay}

We now aim to include the effect of radiative decay from the Rydberg state with rate $\Gamma_\mathrm{ryd}$, described by the dissipator $\mal{D}_\mathrm{dec} \rho =\Gamma_\mathrm{ryd} \suml{k}{}\lt \sigma^-_k \rho \sigma^+_k - \frac{1}{2} \acomm{n_k}{\rho} \rt$. The dynamics of the system are thus described by the von Neumann equation
\begin{equation*}
	\dot{\rho} = \lt \LO + \mal{D}_\mathrm{dec} +  \LI  \rt \rho,
	\label{eq:evol2}
\end{equation*}
where we set the decay rate to be much smaller than the dephasing rate ($\Gamma_\mathrm{ryd}\ll\gamma$), so that we can consider $\mal{D}_\mathrm{dec}$  as a perturbation together with the coherent driving represented by $\LI$.

It is relatively straightforward to prove that
\begin{equation*}
	\comm{P}{\mal{D}_\mathrm{dec}} = 0,
\end{equation*}
so that the analogues of \reff{eq:projevol1} are
\begin{equation*}
	\sysb{l}
		\dot{\chi} =  Q\lt \LO  + \LI+ \mal{D}_\mathrm{dec} \rt \chi + \LI \mu \\[2mm]
		\dot{\mu} =  \mal{D}_\mathrm{dec}  \mu + P\LI \chi.
	\syse
\end{equation*}
By following the same procedure as in \ref{app:general} we can now write
\begin{equation*}
	\dot{\mu} = \mal{D}_\mathrm{dec}   \mu + P\LI \int_0^t \rmd \tau \,\, \rme^{(t-\tau) Q \lt\LO + \LI + \mal{D}_\mathrm{dec}\rt} Q \LI \mu(\tau).
\end{equation*}
For simplicity, we now restrict ourselves to the lowest non-trivial order in both processes, i.e., the decay (expansion in powers of $\Gamma_\mathrm{ryd}$) and the coherent spin-flipping (expansion in powers of $\Omega$). We also consider the order $\Gamma_\mathrm{ryd} \Omega^2$ to be negligible, which allows us to disregard the corrections coming from the substitution $\mu(\tau) \to \mu(t)$ in the expression above. Since the dephasing and decay dissipators commute, our calculation can still hinge on the fact that the long-time behaviour outside of the classical subspace will portray an exponential decay $\approx \rme^{-t \gamma/2}$ (or a faster one). Thus, after the change of variables $\tau \to t-\tau$, we can also bring the upper bound of the integral to infinity, which yields
\begin{equation*}
	\dot{\mu} = \mal{D}_\mathrm{dec}   \mu + P \suml{k}{}  \LI^k \int_0^\infty \rmd \tau \,\, \rme^{\tau Q \LO } Q \LI^k \mu(t).
\end{equation*}
Note that the second addend is simply ${\cal L}^{(2)}$. Thus, we can straightforwardly obtain the expression
\begin{equation*}
	\dot{\mu} = \decay\suml{k}{} Z_k \lqq n_k   \mu \rqq +  \Omega^2 \suml {k}{} \frac{\deph}{\lt \frac{\deph}{2} \rt^2 + h_k(1)^2} Z_k \lqq \mu \rqq,
\end{equation*}
up to first order in $\Gamma_\mathrm{ryd}$ and second order in $\Omega$. This equation is equivalent to \reff{eq:withdec} in the main paper.

\section{Three-level Rydberg atoms in a EIT configuration}\label{app:EIT}

In this Appendix we will give detailed account of the derivation of an effective equation of motion of a system of $N$ driven three-level atoms which display strong interactions between excited states and in the presence of fast decay processes from the intermediate to the ground state, as described in Section \ref{sec:EIT}.

The dynamics of the system is described by the master equation \reff{eq:evol1}, where
\begin{equation*}
	\sysb{l}
		\LO \rho = \mal{D} \rho  \\[2mm]
		\LI \rho = -i\comm{H_0 + H_1}{\rho}
	\syse
\end{equation*}
with
\begin{equation*}
	\sysb{l}
    H_0 = \Delta \suml{k}{} n_k + \ha \suml{k \neq m}{} V_{km} n_k n_m\\[2mm]
    H_1 = \suml{k}{} \lqq  \Omega_p \lt \ket{\downarrow_k} \bra{\leftarrow_k} + \ket{\leftarrow_k} \bra{\downarrow_k}  \rt   +  \Omega_c \lt  \ket{\leftarrow_k} \bra{\uparrow_k} +  \ket{\uparrow_k} \bra{\leftarrow_k}   \rt    \rqq\\[2mm]
    \mal{D} \rho = \Gamma \suml{k}{} \lqq \ket{\downarrow_k} \bra{\leftarrow_k} \rho \ket{\leftarrow_k} \bra{\downarrow_k} - \ha \acomm{\proj{\leftarrow_k}}{\rho}   \rqq.    	
	\syse
\end{equation*}
Note that $H_0$ and $\mal{D}$ act on different subspaces and therefore their actions on the state of the system commute. Therefore, once again, we can include the commutator with $H_0$ in the ``fast'' term $\LO$ while considering only $\mal{D}$ for the determination of the stationary subspace.

While in the previous case the action of the projector $P$ was equivalent to a projection onto the diagonal of the density matrix, here its action is slightly more involved. In order to obtain it, we use the fact that
\begin{equation*}
	 \rme^{ t\mal{D}} =  \rme^{ t \sum_{k}\mal{D}_k} = \prodl{k}{}  \rme^{ t\mal{D}_k}.
\end{equation*}
Each $\mal{D}_k$ non-trivially acts on the $k$-th (three-dimensional) subspace and its action on a generic matrix $A$ can be represented as
\be
	\mal{D}_k A \equiv \mal{D}_k \matb{ccc}  a_{11} & a_{12} & a_{13} \\
							a_{21} & a_{22} & a_{23} \\
							a_{31} & a_{32}  & a_{33}    \mate =
                            \gamma \matb{ccc}  0 & -\frac{a_{12}}{2} & 0 \\
							-\frac{a_{21}}{2} & -a_{22} & -\frac{a_{23}}{2} \\
							0 & -\frac{a_{32}}{2}  & a_{22}    \mate,
							\label{eq:Dk3}
\ee
where we are representing the matrices in the basis ($\ket{\uparrow}, \ket{\leftarrow}, \ket{\downarrow}$). From the relations above, one can extract the action of the projector $P$ as
\begin{equation*}
	P = \prodl{k}{} P_k,\quad\mathrm{with}\quad P_k A = \matb{ccc}  a_{11} & 0 & a_{13} \\
							0 & 0 & 0 \\
							a_{31}  & 0  & a_{33} + a_{22}    \mate,
\end{equation*}
and then define for this case $\mu=P\rho$.

We are now in position to calculate the effective dynamics for $\mu$ up to second order of perturbation in $\LI$ using the results in \ref{app:general}
\begin{eqnarray*}
    {\cal L}^{(1)}  \mu &=& P\LI\mu(t)\\
	{\cal L}^{(2)}  \mu &= & P\LI \int_0^\infty \rmd \tau\,\, \rme^{\tau\LO} Q \LI \mu(t).
\end{eqnarray*}
Let us start by calculating the first order correction. Here, we separate the action of $\LI$ into the one associated to $H_0$ and to $H_1$. The latter can be written as a sum $H_1=\sum_k H_1^k$ and we can therefore restrict our analysis here to a generic ($k$-th) subspace. In particular, $H_1^k$ acts on an element of the kernel of $\mal{D}$ as
\begin{eqnarray}\nonumber
	\comm{H_1^k}{\mu} = \matb{ccc}  0 & \Omega_c & 0 \\
\Omega_c & 0 & \Omega_p \\
0 & \Omega_p & 0    \mate\matb{ccc}  \mu_{\uparrow \uparrow}^{k} & 0 & \mu_{\uparrow \downarrow}^{k} \\
0 & 0 & 0 \\
\mu_{\downarrow \uparrow}^{k} & 0 & \mu_{\downarrow\downarrow}^{k}    \mate - h.c.\\
						=  \matb{ccc}  0 & -\Omega_c \mu_{\uparrow \uparrow}^{k} - \Omega_p \mu_{\uparrow \downarrow}^{k} & 0 \\
						\Omega_c \mu_{\uparrow \uparrow}^{k} + \Omega_p \mu_{\downarrow \uparrow}^{k} & 0 & \Omega_c \mu_{\uparrow \downarrow}^{k} + \Omega_p \mu_{\downarrow\downarrow}^{k} \\
						0 & -\Omega_c \mu_{\downarrow \uparrow}^{k} - \Omega_p \mu_{\downarrow\downarrow}^{k} & 0    \mate,
\label{eq:H1mu}
\end{eqnarray}
which constitutes a matrix orthogonal to the kernel of $\mal{D}$ and, hence, implies that $P\comm{H_1^k}{\mu} = 0$ and $Q\comm{H_1^k}{\mu} = \comm{H_1^k}{\mu}$. Thus, the first order contribution to the effective equation of motion reads
\begin{equation*}
{\cal L}^{(1)}\mu=-i\comm{H_0}{\mu},
\end{equation*}
where we have used that $\comm{\mal{H}_0}{P}=0$ with the shorthand $\mal{H}_0 \bullet = -i\comm{H_0}{\bullet}$.

We now use this knowledge as well to calculate the second order contribution ${\cal L}^{(2)}\mu$. The first thing we realize is that the presence of a projector $Q$ after the application of $\LI$ to $\mu$ leaves only the contribution of $H_1$, which we know already from \reff{eq:H1mu}, as $\comm{\mal{H}_0}{P}=0$. The next step is the application of $\rme^{\tau\LO}=\prod_k\rme^{\tau{\cal D}_k}$ to a matrix of the form \reff{eq:H1mu}. From \reff{eq:Dk3} we can extract that this action amounts simply to a multiplication of the matrix by $\rme^{-\tau\Gamma/2}$. The last step is thus the application of $\LI$ to the matrix \reff{eq:H1mu}. First we realize that, as the interaction Hamiltonian $H_0$ keeps the matrix within the subspace orthogonal to the kernel, its contribution vanishes as $P$ is subsequently applied. Hence, we only need to understand the action of $H_1$, which yields
\begin{eqnarray*}
	&&\comm{H_1^k}{\matb{ccc}  0 & a_{12} & 0 \\
							a_{21} & 0 & a_{23} \\
							0 & a_{32}  & 0    \mate} =  \matb{ccc}  0 & \Omega_c & 0 \\
						\Omega_c & 0 & \Omega_p \\
						0 & \Omega_p & 0    \mate      \matb{ccc}  0 & a_{12} & 0 \\
							a_{21} & 0 & a_{23} \\
							0 & a_{32}  & 0    \mate\! - h.c.  \\
							&& =  \matb{ccc}  \Omega_c \lt a_{21} - a_{12} \rt & 0 & \Omega_c  a_{23} - \Omega_p a_{12}  \\
							0 & \Omega_c \lt a_{12} - a_{21} \rt + \Omega_p \lt a_{32} - a_{23}  \rt & 0 \\
							\Omega_p  a_{21}  -\Omega_c a_{32}  & 0 & \Omega_p  \lt a_{23} - a_{32}  \rt   \mate.
							\label{eq:LIA}
\end{eqnarray*}
Note that this matrix does not generally belong to the kernel of $\mal{D}$, and it is only after applying $P_k$ that one gets
\begin{eqnarray*}
	&&P_k  \matb{ccc}  \Omega_c \lt a_{21} - a_{12} \rt & 0 & \Omega_c  a_{23} - \Omega_p a_{12}  \\
							0 & \Omega_c \lt a_{12} - a_{21} \rt + \Omega_p \lt a_{32} - a_{23}  \rt & 0 \\
							\Omega_p  a_{21}  -\Omega_c a_{32}  & 0 & \Omega_p  \lt a_{23} - a_{32}  \rt   \mate  \\
&&= \matb{ccc}  \Omega_c \lt a_{21} - a_{12} \rt & 0 & \Omega_c  a_{23} - \Omega_p a_{12}  \\
							0 &  0 & 0 \\
							\Omega_p  a_{21}  -\Omega_c a_{32}  & 0 & \Omega_c \lt a_{12} - a_{21} \rt   \mate.
\end{eqnarray*}
Thus, one can obtain now the final form of the second order contribution, which yields
\begin{equation*}
    {\cal L}^{(2)}\mu=\!\frac{2}{\Gamma}\!\sum_k\!\matb{cc}  -2\Omega_c^2 \mu_{\uparrow\uparrow}^k-\Omega_c\Omega_p\epsilon_k & -\lt\Omega_c^2+\Omega_p^2\rt\mu_{\uparrow\downarrow}^k - \Omega_c\Omega_p \delta_k\\
	-\lt\Omega_c^2+\Omega_p^2\rt\mu_{\downarrow\uparrow}^k - \Omega_c\Omega_p \delta_k  & 2\Omega_c^2 \mu_{\uparrow\uparrow}^k+\Omega_c\Omega_p\epsilon_k   \mate\!,
\end{equation*}
with $\epsilon_k=\mu_{\uparrow\downarrow}^k+\mu_{\downarrow\uparrow}^k$ and $\delta_k=\mu_{\downarrow\downarrow}^k+\mu_{\uparrow\uparrow}^k$, and where we have eliminated the intermediate level and hence used a $2\times2$ matrix for the description of the $k$-th atom.

Note that this contribution can be also written out as a purely dissipative Lindblad equation of the form
\begin{equation}
{\cal L}^{(2)}\mu=\suml{k}{} L_k \mu L_k^\dag - \ha \acomm{L_k^\dag L_k}{\mu},
	\label{eq:cal}
\end{equation}
where the jump operators have a non-classical form
\begin{equation*}
	L_k=\frac{2}{\sqrt{\Gamma}}\lt\Omega_c\sigma_-^k+\Omega_p p_k\rt,
\end{equation*}
where $p_k = \proj{\downarrow_k}$ and $\sigma_-^k=\ket{\downarrow_k}\bra{\uparrow_k}$ are spin-$1/2$ operators.

It is worth mentioning that in this simple case one can actually treat the ``classical'' part of the Hamiltonian $H_0$ in a non-perturbative fashion. This can be done employing an interaction representation for $\LO$ and $\LI$:
\begin{equation*}
	\LO \to \wt{\LO} (t) = \rme^{-t \mal{H}_0} \LO \rme^{t \mal{H}_0} \quad \mathrm{and} \quad  \LI \to \wt{\LI} (t) = \rme^{-t \mal{H}_0} \LI \rme^{t \mal{H}_0}.
\end{equation*}
One then finds $\wt{\LO} (t) = \LO$ and $\wt{\LI^k} (t) = -i\comm{\wt{H}_1^k (t)}{ \bullet }$ with
\begin{equation*}
	 \wt{H}_1^k (t) = \rme^{iH_0 t } H_1^k \rme^{-iH_0t} = \matb{ccc}  0 & \rme^{ith_k(1)} \Omega_c & 0 \\
						\rme^{-it h_k(1)}\Omega_c & 0 & \Omega_p \\
						0 & \Omega_p & 0    \mate,
\end{equation*}
where $h_k(1) =  \Delta + \sum_{q\neq k} V_{kq} n_q$ is the same object defined in \reff{eq:hkdef}. The procedure outlined in these appendices can be then carried on in a similar manner; the main difference being that the exponentials $\rme^{t (\LO + \LI)} $ must be replaced by their time-ordered counterparts $T \lqq \rme^{\int_0^t \rmd \tau  \lqq\LO + \wt{\LI}(\tau)\rqq}   \rqq$. At second order in $\LI$ one eventually finds
\begin{eqnarray*}
	\dot{\mu} &= - i \comm{H_0}{\mu} + \suml{k}{} \left\{ \lambda_{cc} \lqq   \sigma_k^- \mal{F}_k   \mu \sigma_k^+ + \sigma_k^- \mu \mal{F}_k^\dag \sigma_k^+  - \mal{F}_k n_k \mu - \mu n_k\mal{F}_k^\dag  \rqq   + \right. \nol[0mm]
	&\left.   + \lambda_{cp} \lqq \sigma_k^- \mal{F}_k \mu p_k + p_k \mu  \sigma_k^+ \mal{F}_k^\dag + \sigma_k^-  \mu p_k + p_k \mu  \sigma_k^+  - \sigma_k^+ \mu - \mu \sigma_k^-  +  \right. \right.   \nol[3mm]
	& \left. \left.   - \sigma_k^- \mal{F}_k \mu  - \mu \sigma_k^+ \mal{F}_k^\dag     \rqq  + \lambda_{pp} \lqq  2p_k\mu p_k - p_k\mu - \mu p_k  \rqq   \right\}
\end{eqnarray*}
where $\lambda_{ij} = 2\Omega_i \Omega_j /\Gamma$ and
\begin{equation*}
	\mal{F}_k = \frac{1}{1 - i \frac{2}{\Gamma} h_k(1)}.
\end{equation*}
We have verified numerically that this expression generally yields negligible corrections to the dynamics with respect to \reff{eq:cal} in the perturbative regime $\Gamma \gg \Omega_{c/p}$.

\section*{References}

\providecommand{\newblock}{}

\end{document}